\documentclass[9pt,twocolumn]{article}

\usepackage{moreverb,url}

\usepackage[colorlinks,bookmarksopen,bookmarksnumbered,citecolor=red,urlcolor=red]{hyperref}
\newcommand{\oursystem}{Anteater}

\newcommand\BibTeX{{\rmfamily B\kern-.05em \textsc{i\kern-.025em b}\kern-.08em
T\kern-.1667em\lower.7ex\hbox{E}\kern-.125emX}}

\usepackage{multirow}
\usepackage{multicol}
\usepackage{tikz}
\usepackage{amsmath}
\usepackage{amsfonts}

\definecolor{myOrange}{rgb}{1,.6,.2}
\definecolor{myBlue}{rgb}{.2,.6,1}
\definecolor{myGreen}{rgb}{.6, .8, 0}

\newcommand{\rf}[1]{{\color{red}}}

\author{Rebecca Faust, Katherine Isaacs, William Z. Bernstein, Michael Sharp, Carlos Scheidegger}%\\Computer Science Department, University of Arizona}
\date{\includegraphics[width=\linewidth]{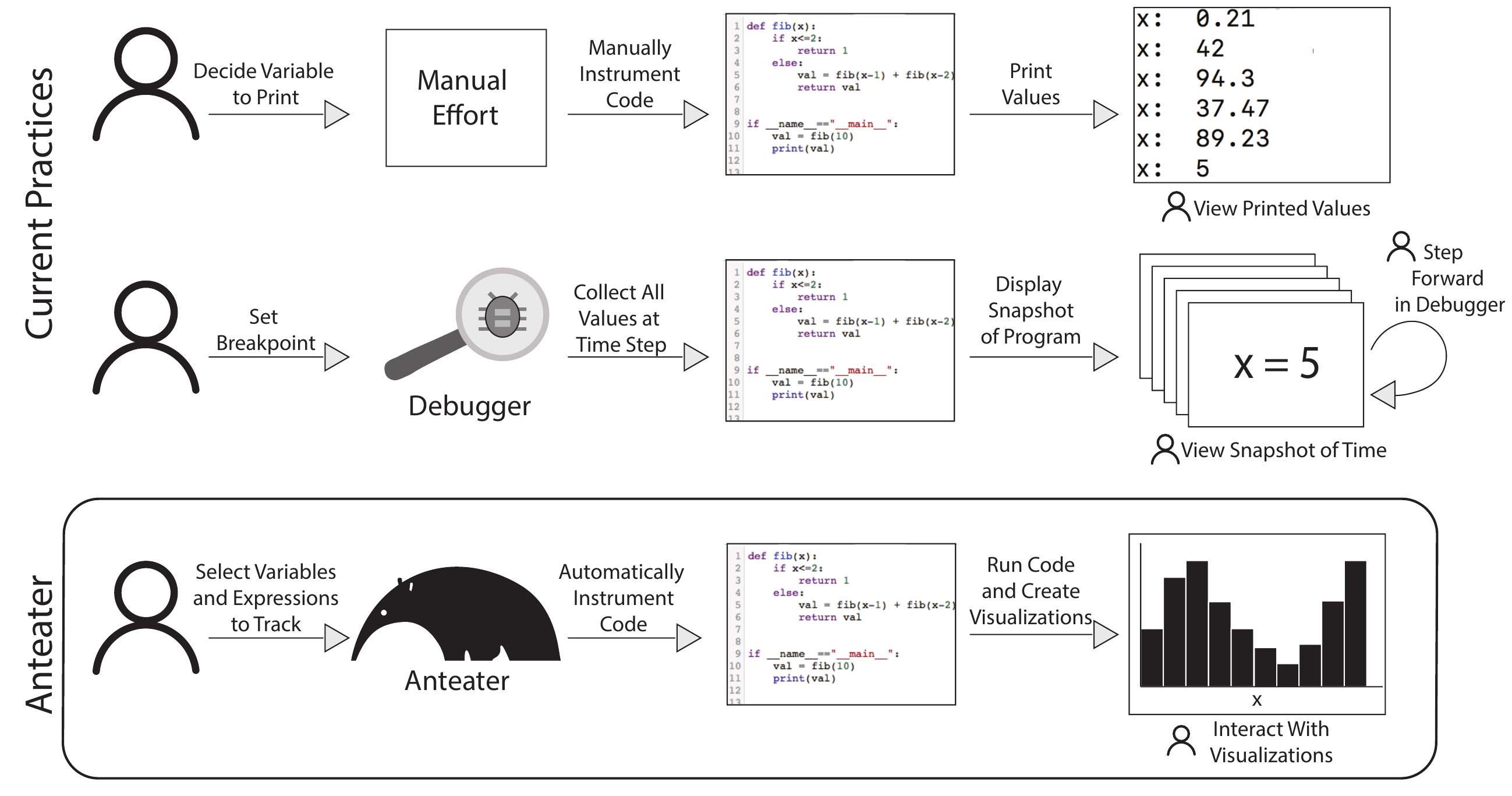}}
\begin{document}

% \runninghead{}

\title{Anteater: Interactive Visualization of Program Execution Values in Context}

% \affiliation{}

% \corrauth{}
% 
% \email{}

\twocolumn[
  \begin{@twocolumnfalse}
  \vspace{-30mm}

    \maketitle
    \vspace{-5mm}
    Figure 1: A programmer investigates a bug in their code.
One common practice (top row) is to instrument the program manually to collect suspicious variables (here, $x$), and print their values.
Manual instrumentation, however, is itself repetitive and error-prone.
Another common practice (second row) is to use a debugger to stop the execution of the program and view each individual value assignment of $x$, providing a precise, but narrow, one-at-a-time view of the values.
\oursystem{} (bottom row) automatically instruments the code to track variables along with the context of their execution.
It presents the programmer with interactive visualizations providing a global view of values, enabling easy detection of erroneous values as well as interactions that narrow down the views to specific values. 
\vspace{-2mm}
\paragraph{Abstract - }Debugging is famously one the hardest parts in programming.
  In this paper, we tackle the question: what does a debugging environment look like when we take interactive visualization as a central design principle?
  We introduce \oursystem{}, an interactive visualization system for tracing and exploring the execution of Python programs.
  Existing systems often have visualization components built on top of an existing infrastructure.
  In contrast, \oursystem{}'s organization of trace data enables an intermediate representation which can be leveraged to automatically synthesize a variety of visualizations and interactions.
  These interactive visualizations help with tasks such as discovering important structures in the execution and understanding and debugging unexpected behaviors.
  To assess the utility of \oursystem{}, we conducted a participant study where programmers completed tasks on their own python programs using \oursystem{}.
  Finally, we discuss limitations and where further research is needed.
\vspace{5mm}
  \end{@twocolumnfalse}
]
% \keywords{Interactive Visualization, Program Traces}

% \maketitle

% \begin{figure*}
%     \centering
%   \includegraphics[width=.8\linewidth]{teaserWider.pdf}
%   \caption{A programmer investigates a bug in their code.
% One common practice (top row) is to instrument the program manually to collect suspicious variables (here, $x$), and print their values.
% Manual instrumentation, however, is itself repetitive and error-prone.
% Another common practice (second row) is to use a debugger to stop the execution of the program and view each individual value assignment of $x$, providing a precise, but narrow, one-at-a-time view of the values.
% \oursystem{} (bottom row) automatically instruments the code to track variables along with the context of their execution.
% It presents the programmer with interactive visualizations providing a global view of values, enabling easy detection of erroneous values as well as interactions that narrow down the views to specific values. }
% 	\label{fig:teaser}
% \end{figure*}

\section{Introduction}

% In this paper, we tackle the following questions: 
% \textbf{when we take interactive visualization principles as a driving concern, what novel designs are possible for debugging systems?} \textbf{Can interactive visualization offer unique benefits for program debugging?}\rf{Reviewers have issues with us not directly addressing these questions}
Debugging and understanding program behavior is notoriously one of the most burdensome aspects of programming. It often requires programmers to trace through the execution steps and values of their program. However, most tools require people to build mental traces of their programs through the serial inspection of program values. 
Current practices often involve stepping through debuggers, inserting logging statements, or searching through source code, either manually or with a code browsing tool~\cite{latoza2010developers}. 
% Debugging is time consuming. Current practices often involve stepping through debuggers, logging statements, or searching through source code, either manually or with a code browsing tool~\cite{latoza2010developers}. 

Additionally, traditional debuggers require programmers to set breakpoints at which they inspect the program state, stepping through its line-by-line operation.  Tiarks et al.~\cite{tiarks2012challenges} observed that programmers experience difficulties in choosing breakpoint locations, often forgetting analysis details while navigating the code. Furthermore, traditional debuggers only present one view of the program: the whole program state at a single step in time.  While this view has its uses in debugging, it does not help with bugs that present themselves over time (i.e. bugs where viewing a single instance of a variable is insufficient for detecting the bug, see Gradient Descent usage scenario). To detect those bugs, programmers must serially step through the values to build a mental image of their behavior. 

However, this method of incrementally inspecting values to build an internal mental image of data directly contrasts the fundamental principles of data visualization. Consider the traditional value proposition of data visualization. 
Visualization practitioners now have a well-defined set of principles to drive the design, development, and testing of interactive visualization software~\cite{bertin1983semiology,carpendale2003considering,shneiderman2003eyes}.
In contrast to inspecting datasets serially, one element at a time, well-designed visual encodings
can provide richer, faster, and more global views of potentially important patterns.

Because traditional debugging methods only provide serial views of program data, they suffer from the same fundamental problem associated with the serial inspection of data.
The widely used ``Visual Information Seeking Matnra'', as presented by Shneidermann~\cite{shneiderman2003eyes}, states ``overview first, zoom and filter, then details-on-demand''. We have seen numerous successful applications of this mantra to data analysis problems. However, we have yet to see this applied in a debugging context where serial inspection of data remains as the primary analysis method. We therefore see a need for an exploratory debugging solution that provides more effective global views of values, providing debugging the same set of affordances that interactive visualization provides
to exploratory data analysis.

Consider the following debugging scenario.  Programmer Patty has a bug in her code. Her program returns a value that seems unreasonable. She believes that the bug is occurring in a specific loop but cannot identify the root cause.  Using a typical debugger, she sets a breakpoint at the beginning of the loop and runs the debugger.  When the debugger reaches the breakpoint, she inspects the program values and takes a few steps through execution but does not yet see the bug. Patty continues the program until it hits the breakpoint again at the next iteration, repeating this process.  She continues to step through each iteration of the loop but has little success in finding the bug.  

After several iterations, Patty gives up on using the debugger and modifies the code with print statements.  She prints the variable she believes causes the bug and runs the program. Patty scans through the printed values, trying to find any erroneous values, but her loop has many iterations and she quickly gets lost in the print statements.  

Her next idea is to write the values to a file and plot them.  Patty first alters her source code to write the values to a file.  She then writes a script that reads the file and plots the values.  Now she sees the behavior of every instance of the value and pinpoint the incorrect values. With this information, Patty returns to the debugger and stops the program when it reaches the iteration containing incorrect values to find the root cause. 

The scenario described above encompasses the typical ways programmers debug their programs~\cite{tiarks2012challenges}.  While not every bug requires all of these methods, programmers typically use more than one of them.  The fact that many programmers use a combination of independent debugging-methods when fixing their programs prompts the question: can we design a better debugger that 1) reduces the amount of manual instrumentation required, 2) gives the users greater control over the values they see, and 3) provides them with a visualization option automatically?  While various debugging tools address aspects of these problems, no existing debugger comprehensively addresses all of them. 
% There exist tools that add visualizations on top of existing debuggers. However, many of them still operate within a snapshot of the program. Rather than visualizing the global behaviors of values, such tools help visualize more complex objects at a given timestep.  
% In contrast, we propose \oursystem{} alone addresses the above design question by combining value visualizations, execution structure visualizations, and whole time data views.

In response to these questions, we present \oursystem{}, a system for debugging and understanding programs designed with principles of interactive visualization as a driving concern. 
% We designed \oursystem{} to address these problems in a more comprehensive way by using the principles of interactive visualization. 
We applied the framework for visual design as described by Munzner~\cite{munzner2014visualization} to create a debugging system from a visualization perspective. Fig.1 gives an overview of how \oursystem{} compares to standard debugging practices.
In taking a visualization-first approach, \oursystem{} provides more informative overviews of a program's behavior while supporting interaction to dig deeper into the details of the execution.  Rather than showing the whole state at a single step in time, it shows a single variable over the entirety of the execution.  
\oursystem{} aims to reduce the effort required from a user by 1) automatically instrumenting programs to collect the values they want to inspect and 2) allowing them to browse values of interest easily throughout the entire execution, without resorting to a step-through debugger.

%  Rather than presenting users a snapshot of everything at a single timestep, we present them with global views of targeted variables of interest with interactions to narrow down to specific values.  In doing so, users can more easily discover patterns within the values they deem important. We pair these global views with a visualization of the execution structure that allows users to maintain context with the execution and inspect execution information surrounding the values. Fig.\ \ref{fig:fibOverview} presents an overview of \oursystem{}. 

 If Programmer Patty had been using \oursystem{}, she could have easily set \oursystem{} to track the value she believed to be raising issues along with any other values that she believed to be potential roots of causation. \oursystem{} would then trace her program and provide her with visualizations to help her identify the iterations where the value was incorrect. Patty could then filter down the execution tree to those iterations and inspect the rest of the values she tracked. With \oursystem{}, Patty completes all of her debugging in one place using only a few interactions and requiring no manual instrumentation.

% Another common problem users face when programming is that of making sense of unfamiliar code~\cite{tiarks2012challenges}.
% Whether programs are pulled from the internet, handed off to others by a collaborator, or existing in a just-joined project, sifting through code to understand its execution and determining where to start working is challenging.
% Something as simple as identifying the dependencies of a variable can become a significant burden when the program is large.
% There is a need for facilitating a deeper understanding of the structure of a program's execution to assist programmers in exploring how functions, variables, and values depend on one another.

\begin{figure*}
    \centering
    \includegraphics[width=.8\linewidth]{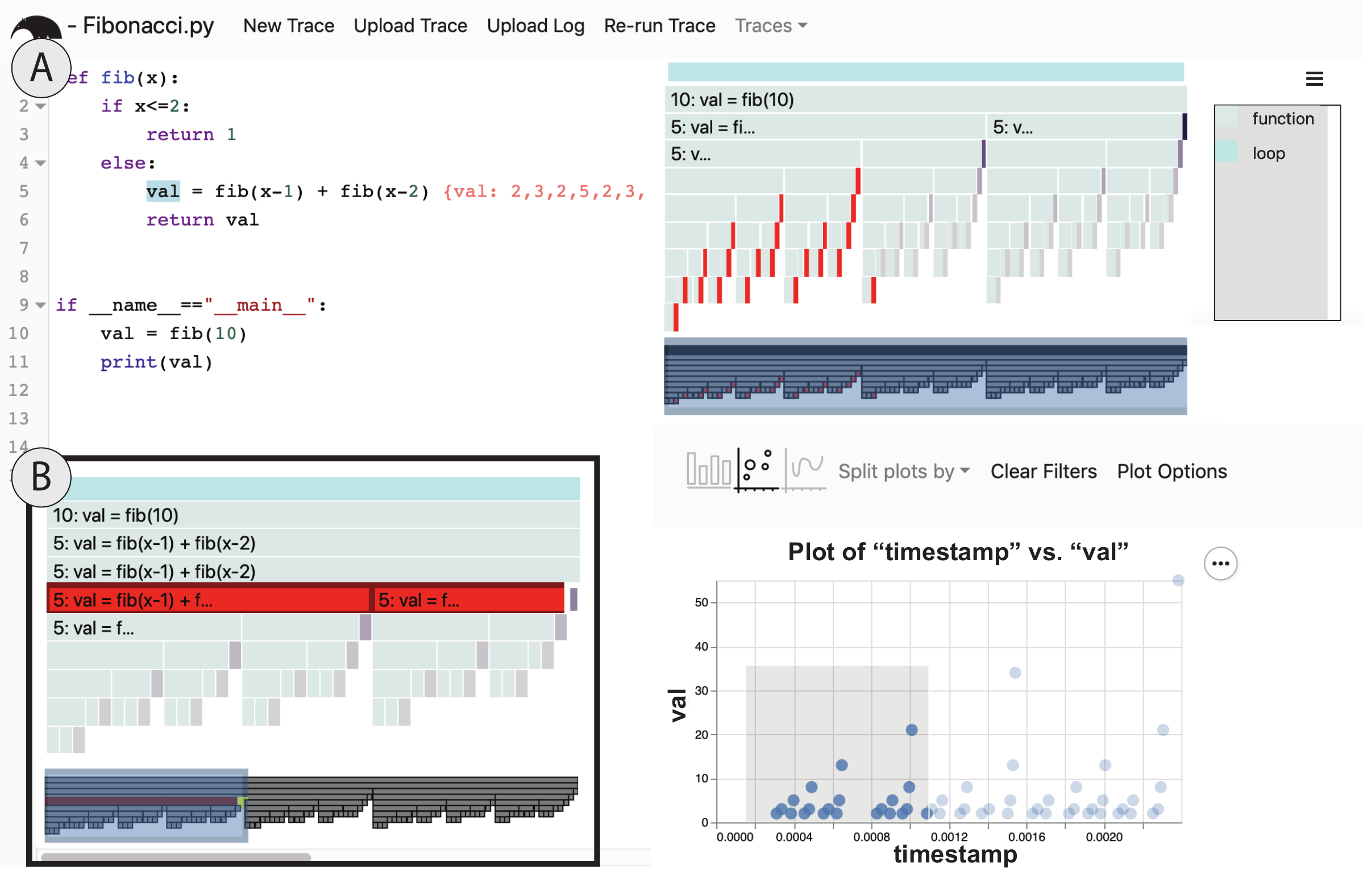}
    \caption{An overview of the \oursystem{} UI on a recursive Fibonacci program, tracking the variable ``val".  (A) shows the UI presented by \oursystem{} (not including (B)).  The generalized context tree (GCT), or icicle plot, shown on the top right side of (A), shows the structure of the execution trace. The teal blocks represent function calls while the varying shades of purple represent the value of ``val" at that instance. We can see the recursive calling structure of the Fibonacci function and can easily identify where it is repeating work. The plot currently shows a scatterplot view of the variable ``val" over time. Brushing over the scatterplot highlights the corresponding instances in the GCT (the red blocks shown in the GCT on the right side of (A)) and the context bar. The scatterplot shows repetitive patterns that indicate that Fibonacci is doing redundant work.  (B) shows a second view of the GCT (inset into the image of the main UI) after we've clicked on a block in the tree which caused its dependencies to be highlighted in red. This shows that the selected block (on the far right of the fifth row in the GCT in (B)), representing an instance of ``val", depends on the prior two calls to the Fibonacci function (shown by the two blocks highlighted in red). }
    \label{fig:fibOverview}
\end{figure*}

In this paper, we present a prototype implementation in Python that traces a Python program to capture not only the execution structure but also values of interest in context of the execution. 
\oursystem{} then presents this trace to the user through interactive visualizations. Fig.~\ref{fig:fibOverview} presents an overview of the visualizations provided by \oursystem{}.
In summary, this paper contributes (i) a goals-and-tasks analysis~\cite{lam2018bridging} of the typical practice of program debugging, (ii) a description and prototype implementation of \oursystem{} in Python, aimed at providing interactive-visualization support to program debugging and understanding, (iii) a paired analytics evaluation of the prototype and its analysis, (iv) a comparative evaluation of \oursystem{} and a traditional IDE debugger, and (iv) usage scenarios of real-world programs that show how our system compares to existing approaches.

\begin{table*}[htbp]
    \centering
    \small
    \begin{tabular}{|c||c|c|c|c||c|c|c|c|c|}
        \hline
        \multirow{5}{1.8cm}{\centering Tool}&
        \multicolumn{4}{c||}{\tikz\draw[myGreen,fill=myGreen] (0,0) circle (.5ex); Supported Views} & \multicolumn{4}{c|}{ \tikz\draw[myGreen,fill=myGreen] (0,0) circle (.5ex); Features}\\
        \cline{2-9}
         & \multirow{4}{1.1cm}{\centering Single Variable, Single Time} &  \multirow{4}{.9cm}{\centering Whole State, Single Time }&   \multirow{4}{1.1cm}{\centering \tikz\draw[myBlue,fill=myBlue] (0,0) circle (.5ex); Single Variable, Whole Time }&  \multirow{4}{.9cm}{\centering Whole State, Whole Time} &  \multirow{4}{1.35cm}{\centering Breakpoint/ Step-through} &  \multirow{4}{1.55cm}{\centering Variable Visualization(s) (*interactive)} &  \multirow{4}{1.2cm}{\centering  \tikz\draw[myOrange,fill=myOrange] (0,0) circle (.5ex); Execution Structure Visualiza-tion} & \multirow{4}{1.8cm}{\centering  Limitations} \\ 

         &&&&&&&&\\
         &&&&&&&&\\
         &&&&&&&&\\
         &&&&&&&&\\
         \hline

        \multirow{4}{*}{\centering Anteater} &
        \multirow{4}{*}{\centering \checkmark} &&
        \multirow{4}{*}{\centering \checkmark} &&& \multirow{4}{*}{\centering \checkmark$^*$}& \multirow{4}{*}{\centering \checkmark}& \multirow{4}{1.5cm}{\centering Medium Scale Python Programs}\\
        &&&&&&&&\\
        &&&&&&&&\\
        &&&&&&&&\\
        \hline     
        \hline

        \multirow{4}{1.3cm}{\centering Omnicode ~\cite{kang2017omnicode}} & 
        \multirow{4}{*}{\centering \checkmark} & \multirow{4}{*}{\centering \checkmark} & \multirow{4}{*}{\centering \checkmark} & \multirow{4}{*}{\centering \checkmark} & & \multirow{4}{*}{\centering \checkmark$^*$} & & \multirow{4}{1.5cm}{\centering Python Programs w/ $< 10 $ variables}\\
          
        &&&&&&&&\\
        &&&&&&&&\\
        &&&&&&&&\\
        \hline
      
        \multirow{4}{*}{\centering GDB} &
        \multirow{4}{*}{\centering \checkmark} & \multirow{4}{*}{\centering \checkmark} &&&\multirow{4}{*}{\centering \checkmark} &&&\\
        &&&&&&&&\\
        &&&&&&&&\\
        &&&&&&&&\\
        \hline
        
        \multirow{4}{1.2cm}{\centering Print Statements} &
        \multirow{4}{*}{\centering \checkmark}  & & \multirow{4}{*}{\centering \checkmark} & &  &&&\\
        &&&&&&&&\\
        &&&&&&&&\\
        &&&&&&&&\\
        \hline
     
        \multirow{5}{1.5cm}{\centering Traditional Visual Debuggers ~\cite{mirur,Gestwicki2005JIVE,Reiss2014CodeBubbles,rozenberg2014templated,Cheng2016}} &
        \multirow{4}{*}{\centering \checkmark} & \multirow{4}{*}{\centering \checkmark} & && \multirow{4}{*}{\centering \checkmark}& \multirow{4}{*}{\centering \checkmark$^{(*~\cite{mirur})}$ }&\multirow{5}{1.2cm}{\centering \checkmark- ~\cite{Gestwicki2005JIVE,Reiss2014CodeBubbles}} & \multirow{4}{1.5cm}{\centering ~\cite{mirur,Gestwicki2005JIVE,rozenberg2014templated,Cheng2016} - Eclipse Plugin}\\
        &&&&&&&&\\
        &&&&&&&&\\
        &&&&&&&&\\
        &&&&&&&&\\
        \hline
     
        \multirow{4}{1.7cm}{\centering Memory Visualization ~\cite{Aftandilian2010,litvinov2017tool,Sundararaman2008HDPV}} & 
        \multirow{4}{*}{\centering \checkmark} & \multirow{4}{*}{\centering \checkmark} & && \multirow{4}{*}{\centering \checkmark-~\cite{Sundararaman2008HDPV}}& \multirow{4}{*}{\centering \checkmark$^{(*~\cite{Aftandilian2010,Sundararaman2008HDPV})}$} & & \multirow{4}{1.5cm}{\centering ~\cite{litvinov2017tool} - Eclipse Plugin}\\
        &&&&&&&&\\
        &&&&&&&&\\
        &&&&&&&&\\
        \hline
     
        \multirow{6}{1.7cm}{\centering Trace Visualization  ~\cite{Bezemer2015FlameGraph,gralka2017visual,Karran2013SyncTrace,Renieris1999ALMOST,Trumper2010}} & 
        & & &\multirow{5}{*}{\centering \checkmark} & &  &\multirow{5}{*}{\centering \checkmark}& \multirow{6}{1.8cm}{\centering Don't track values,  ~\cite{gralka2017visual,Karran2013SyncTrace,Renieris1999ALMOST,Trumper2010} - Limited Scale - Tracing Overhead}\\
        &&&&&&&&\\
        &&&&&&&&\\
        &&&&&&&&\\
        &&&&&&&&\\
        &&&&&&&&\\
        \hline
     
        \multirow{6}{1.8cm}{\centering Debuggers with Global Visualizations ~\cite{alsallakh2012visual,beck2013visual,Burg2013,hoffswell2018augmenting,Schulz2016}} &
        \multirow{4}{*}{\centering \checkmark} &  &\multirow{4}{*}{\centering \checkmark} &\multirow{4}{*}{\centering \checkmark-~\cite{beck2013visual}}& \multirow{4}{*}{\centering \checkmark}& \multirow{4}{*}{\centering \checkmark$^{(*~\cite{alsallakh2012visual,Burg2013,hoffswell2018augmenting,Schulz2016})}$} & & \multirow{4}{1.5cm}{\centering ~\cite{alsallakh2012visual,beck2013visual} - Eclipse Plugin, }\\
        &&&&&&&&\\
        &&&&&&&&\\
        &&&&&&&&\\
        &&&&&&&&
        ~\cite{Burg2013} - Web, \\
        &&&&&&&&~\cite{hoffswell2018augmenting} - Vega\\
        \hline
        
        \multirow{4}{1.2cm}{\centering  ~\cite{hoffswell2016visual}} & 
        \multirow{4}{*}{\centering \checkmark} & \multirow{4}{*}{\centering \checkmark} &\multirow{4}{*}{\centering \checkmark} &\multirow{4}{*}{\centering \checkmark}&  & \multirow{4}{*}{\centering \checkmark *} & & \multirow{4}{1.5cm}{\centering Vega Specifications }\\
        &&&&&&&&\\
        &&&&&&&&\\
        &&&&&&&&\\
        \hline

\end{tabular}

    \caption{A comparison of Anteater with existing work in debugging visualizations. This table contains most references from the Related work section with the exception of ~\cite{Do2018VisuFlow, Lieberman1995ZStep95, Ohmann2016Perfume, Socha1988Voyeur} which did not fit into the above categories.  In this table, "single time" refers to a single instance at a specific point in the execution whereas "whole time" refers to every instance throughout the entire execution. The colored circles correspond to views and features that support the goals defined later in this paper. Note, because of the generality of {\color{myGreen} G3}, all systems aim to support this goal in some capacity and as a result, all features and views support it in some way.  When a cell specifies specific references (e.g. \checkmark-~\cite{beck2013visual} or ~\cite{hoffswell2018augmenting} - Vega ) this means that only those references have the corresponding view or feature.  
  }
    \label{tab:comparisons}
\end{table*}

\section{Related Work}

\paragraph{Literature Search}
We compare \oursystem{} to work we have found in software engineering, user interface design, information visualization, and visual analytics.
Specifically, we have searched the last 25 years of work related to visual debugging in the following venues: ACM ICSE, ACM CHI, ACM UIST, IEEE VIS, and the SoftVis symposium. The field of software visualization is large and we cannot hope to add every possible reference; we recommend both textbooks from Diehl and Stasko as starting points into the literature~\cite{diehl2007software,stasko1998software}.  Figure 1 gives a general overview of how ~\oursystem{} compares to common debugging methods and table~\ref{tab:comparisons} gives an overview of how \oursystem{} compares to the relevant existing work discussed in the this section.

\paragraph{Visual Debugging} Many attempts have been made to leverage visualization principles to augment the debugging process.
Some efforts add visualization options to breakpoint and step-through debuggers~\cite{Cheng2016,Lieberman1995ZStep95,Do2018VisuFlow,Reiss2014CodeBubbles,mirur,rozenberg2014templated,litvinov2017tool}.
Traditional visual debuggers typically provide visualization views of variables at a specific instance in time, much like traditional debuggers. Several of these tools add visualizations of objects to a traditional debugger~\cite{mirur,rozenberg2014templated,Cheng2016}.
Others provide visualizations to show task-specific information about the execution, such as an overview of the heap and stack~\cite{Aftandilian2010, litvinov2017tool} the impact of resource utilization on control flow~\cite{Ohmann2016Perfume}, object mutation~\cite{Schulz2016}, or run-time state and data structures of the program~\cite{Sundararaman2008HDPV}.

Generally, these tools present localized views that describe one particular state of the execution.
Some tools provide additional context by allowing back-stepping in the debugger or providing a history of the execution~\cite{Reiss2014CodeBubbles,Gestwicki2005JIVE,Lieberman1995ZStep95}.  In addition, some tools provide global views to show the behavior of values over the entire execution. 
Aftandilian et al.~\cite{Aftandilian2010} give a global view of the heap by taking snapshots throughout the program.
Schulz et al.~\cite{Schulz2016} provide a global view of object mutations; if the object is numeric, the global view shows the value behavior throughout the execution.
Some tools give global views of value behaviors by introducing sparklines next to the line of source code defining the value~\cite{beck2013visual,hoffswell2018augmenting}.
In contrast, \oursystem{} displays global views that take the execution context into account.
As we show in our evaluation, this perspective can be particularly helpful in debugging scenarios.

Hoffswell et al.~\cite{hoffswell2016visual} and Burg et al.~\cite{Burg2013} describe systems for visually debugging user interactions, one on Vega specifications and the other on web applications in general.  Similar to Anteater, both systems recognize the importance of recording program behavior and providing global views of data to understand the inner-workings of a program. They differ from Anteater in their focus on debugging interactions with an application rather than the execution of a program.

Alsallakh et al.~\cite{alsallakh2012visual} created an Eclipse plugin that tracks specific tracepoints (equivalent to a breakpoint in a debugger) throughout a program's execution.
Watchpoints can also be added to a field on which the tracer will track assignments.
The tool provides global views of tracepoint instances through line charts where interactions provide additional information about the program at that point and watchpoints through a step chart of the values over time.
While the plugin's goals closely relate to those of our prototype, Anteater stands apart for two reasons.
First, \oursystem{} traces all calls and loops, rather than user-defined tracepoints, along with the values desired by the user.
Second, \oursystem{} presents all this information in a trace visualization with corresponding plots of the tracked values.
This information can provide the context necessary to better understand why variables take on certain values.

The most similar tool to \oursystem{} is Kang et al.'s~\cite{kang2017omnicode} Omnicode.
Omnicode provides run-time visualizations of program states, designed to aid novice users in building mental models about programs.
Crucially, Omnicode visualizes values in a live coding environment which updates in real time.
The primary visualization provided is a scatterplot matrix displaying plots for each variable over all execution steps.
While Omnicode and \oursystem{} have much in common, they were designed for different audiences (novices vs.\ general programmers) and thus support different types of programs.
We compare Omnicode and \oursystem{} directly in the Discussion section.

% \paragraph{Tracing}
% Historically, program traces have captured performance information for a program's execution.  They typically track function calls and time spent within a function.  Visualizations are then created to present this information to users.  However, function calls and execution time only solve a subset of bugs. 

\paragraph{Trace Visualization} 
Trace visualizations are often applied in support of understanding parallel programs~\cite{Socha1988Voyeur,Karran2013SyncTrace,Trumper2010}.
Often, trace visualizations leverage icicle plots and flame graphs as the primary visual representation~\cite{Renieris1999ALMOST,Karran2013SyncTrace,gralka2017visual,Bezemer2015FlameGraph,Trumper2010}.
\oursystem{} uses a visual encoding reminiscent of icicle plots and flame graphs in our plots of the execution trace, which we will call the generalized context tree (GCT), after Boehme et al.~\cite{Boehme2016}
However, \oursystem{} differs in its definition of \textit{trace}.
While these previous traces capture the calling structure of the execution, \oursystem{} extends this to capture values of marked variables and expressions, as well as loop behaviors.
This extension provides users with additional context for how values are reached; see Evaluation for a discussion of their utility.

\section{Characterization of \phantom{wordwo} \oursystem{}'s Design}
This section describes the visualization framework used to characterize the problem of program debugging from a visualization perspective. This perspective drove the design \oursystem{}. Additionally, it uses the taxonomy presented by Maletic et al~\cite{maletic2002task} to characterize the system design of \oursystem{}.

\subsection{A Visualization Perspective on \phantom{to} Program Debugging }
In this section, we use the framework for visual design described by Munzner~\cite{munzner2014visualization}  to characterize the problem of program debugging from a visualization perspective. This was the driving perspective used to create \oursystem{}. The framework consists of 3 parts: (1) what - the data abstraction, (2) why - the task abstraction, and (3) how - the actual visualization design. This section describes how debugging maps to this framework, with the following three sections describing in detail how \oursystem{} applies this perspective.

\paragraph{What - Data Abstraction:}
First, we need to understand the data involved in program debugging.  As a program executes, it inherently creates a collection of data. This data includes items such as the values assigned to every variable, the value of parameters passed into function calls, the structure of the execution (e.g. calls and loops), time spent in each part of the program, etc. This data naturally maps to the data types outlined in the framework.  We will focus on a subset of the data generated from sequential programs: the structure of the execution and the values assigned to variables. These two forms of data correspond to two data types outlined in Munzner's framework.

The first data type is a tree. A sequential programs naturally executes in a hierarchical tree structure: the root of the tree represents the entry point into the program, nodes represent execution steps (e.g. functions and loops), and the parent/child relationship signifies that the child was executed within the parent instruction (e.g. within a function call or loop iteration).
% : each line must fully execute before the following line can execute.  If a given instruction is a function call, it must execute all internal code before it the program can return and execute the next instruction. Similarly, each iteration of a loop must fully execute its internal instructions before it can move on (either to the next iteration or, if it broke out of the loop, to the next sequential instruction). This hierarchy translates into a tree. 
The tree creates a node every time the program enters a function call or an iteration of a loop and it creates an edge between each node and the parent function or loop that contains its instruction. In Munzner's framework, this corresponds to the data types \textit{node} and \textit{link} and the dataset type \textit{network/tree}. Additionally, each function call and loop contains additional \textit{attributes}, such as the source code line that corresponds to its instruction, the name of the function, the value of the iterator for the loop, etc.   

The second data type is a table. The values of program variables naturally organize into tables.  Each instance of a variable is a data \textit{item} (a row in the table) that contains several \textit{attributes} that describe it (the columns in a table). To construct these tables, a program must create a record every time the program assigns to a variable. The attributes associated with a variables assignment include the line at which the assignment occurred, the node in the execution tree that contains that assignment, the actual value of the variable at that instance, etc. This clearly corresponds to \textit{table} dataset in the framework, with the \textit{item} and \textit{attribute} data types describing the entries in the table. 

\oursystem{} uses this data abstraction to create  a visual representation of a program. We describe the generation of this data with \oursystem{} in more detail in the section titled ``Tracing Infrastructure and Data Organization''.

\paragraph{Why - Task Abstraction}
Now that we understand the data abstraction, we need to understand how the data analysis actions and targets outlined in the framework map to the domain of program debugging. 

The high-level goal of debugging is to discover the source of unexpected or erroneous program behavior.  This behavior could either stem from misbehavior in the execution structure (e.g. a function not being called as expected) or misbehavior in the variable values (e.g. an incorrect calculation), or both.  When debugging, programmers often inspect the programs data to generate a hypothesis about why the program is misbehaving or to validate an existing hypothesis about a bug. In Munzner's framework, this goal falls into the \textit{consume} action of the \textbf{analyze} category. Additionally, this goal corresponds to the overarching aim of \oursystem{}.

The framework allows us to separate the high-level goal of discovering unexpected program behavior into 4 mid-level actions that correspond to the programmers prior knowledge of the bug: \textit{lookup}, \textit{locate}, \textit{browse}, and \textit{explore}. These actions fall into the \textbf{search} category of Munzner's framework. First, a programmer may know precisely what to look for and where to look for it (corresponding to the \textit{lookup} action). For example, if through prior debugging efforts they identified and corrected a calculation error, they may then re-execute the program to lookup the new value to ensure that it is correct. In this case, they know exactly what they are looking for and where to find it in the program data. 

Second, a programmer may know what the bug is but not where it is occurring (corresponding to the \textit{locate} action). For example, if a programmer knows that their program is producing an erroneous output value, they know that somewhere in the execution an erroneous value is assigned to the variable but they don't know precisely where. 

Third, a programmer may know the general location of a bug, but not precisely what is causing it (corresponding to the \textit{browse} action). For example, a program deviates from the expected execution path at line $x$ but the programmer cannot immediately see what causes the deviation. They must browse the program data around this point to understand the behavior of the program at that point.

Fourth, a programmer may not know where the bug is or what is causing it (corresponding to the \textit{explore} action). For example, a program finishes running but does not return from the expected point in the program. The programmer must explore the program data to locate where the program returns from and why it returns from this point instead of the expected point. 

At the lowest level of action exist the \textbf{query} actions. These actions correspond to specific ways in which a programmer might query their program data for a debugging tasks. While performing a \textit{lookup} or \textit{locate} action where the programmer knows what variable or function call causes a bug, the may \textit{identify} the specific instance of that call or variable to inspect all of the information collected about that instance. In contrast, when performing a \textit{browse} or \textit{explore} action, programmers want to \textit{identify} areas in the program data that deviate from their expectations. A programmer may also want to \textit{compare} the values of two variables to understand or verify an expected relationship between them.  Last, for observing trends or patterns and identifying potential areas of erroneous behavior in variables or the execution structure, programmers may want to \textit{summarize} the data with global views of the data. 

The targets of these actions may be trends, outliers, or features of variable values or execution structure that highlight the misbehavior. They may also be correlations between variables that are perceived to be related or the distribution of a single variable.  When identifying unexpected execution structure, the target  may be the topology of the execution tree and paths through the tree that correspond to the execution stack of the program.

\oursystem{} provides interactive views of both the execution structure and variable value data that allow people to perform these actions on program data. Global views of the program value allow people to \textit{browse}, \textit{explore}, and \textit{summarize} the data. Interactions on the global views allow people to narrow their view to perform \textit{lookup} and \textit{locate} actions. The ability to plot multiple variables on a single plot allows people to \textit{compare} variables.

\paragraph{How - Visual Design }
With the data abstraction and task abstraction defined, all that remains is creating visualizations of the data that facilitate the specified tasks. While there exist numerous options for creating visualizations of this data, we will focus solely on those supported by \oursystem{}. We will not go into detail about the visual design here but will give a high level description of how \oursystem{}'s visualizations map to the framework. A full description of the visual design can be found in the section ``Anteater's Visualization Design''.  The framework breaks up visualizations into four classes: \textbf{encode}, \textbf{manipulate}, \textbf{facet}, and \textbf{reduce}.

\oursystem{} \textbf{encodes} the data using color and arrangement. For variable values, depending on the type of variable, \oursystem{} arranges the data tables into histograms, barplots, scatterplots or parallel coordinates. \oursystem{} arranges the execution tree into an icicle plot to illustrate the hierarchical structure of the execution and creates a color map to signfiy the type (function call, loop, etc.) of each block in the icicle plot. 

\oursystem{} allows programmers to \textbf{manipulate} the data through selections on the plots and execution tree. This enables them to connect the two views and inspect specific values in the visualization.  Programmers then can \textbf{reduce} the data by filtering their selections to exclude irrelevant information. 

Last, \oursystem{} allows programmers to \textbf{facet} their data by partitioning it using a shared structure in the execution (such as a repeated function call or loop iterations) or related values from the program (such as a related boolean variable). 

\begin{figure*}
    \centering
    \includegraphics[width=\linewidth]{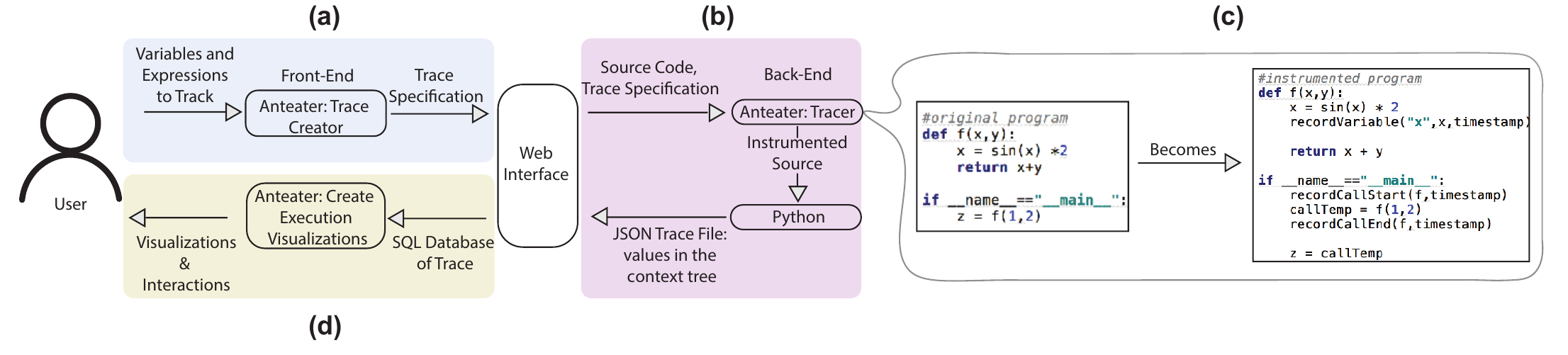}
    \caption{An overview of the \oursystem{} system. In (a),  a user chooses variables and expressions to track using the \oursystem{} interface. This defines the trace specification. Then, \oursystem{} sends the trace specification through the web interface to the python backend, along with the source code.  Next, in (b), the \oursystem{} tracer instruments the source code to collect execution information along with the specified values.  (c) shows a simplified version of this instrumentation. After the code is instrumented, \oursystem{} runs the program using python to create the program trace.  This trace is passed back through the web interface to the \oursystem{} front end where (in (d)) it is visualized and presented to the user. }
    \label{fig:sysOVerview}
\end{figure*}

\subsection{Characterizing \oursystem{}'s System Design}
Several taxonomies exists for characterizing program visualizations~\cite{stasko1992understanding, roman1992program, price1993principled,maletic2002task,sensalire2008classifying}. While any of the taxonomies can apply to \oursystem{}, we use the taxonomy from Maletic et al.~\cite{maletic2002task} to describe it because we believe that it best characterizes \oursystem{} with respect to the systems goals. This taxonomy breaks program visualizations into 5 dimensions: Tasks, Audience, Target, Representation, and Medium. We discuss each of these dimensions individually in the remainder of this section.

\paragraph{Tasks} 
The task dimension, as specified by Maletic et al., defines why the visualization is needed. Most standard debugging tools and methods lack support for global visual representations of the data internal to programs. They rely on serial approaches of inspecting a single instance of the data at a time. However, serial inspection of raw data tasks people with the significant mental burden of building an internal representation of an entire dataset~\cite{munzner2014visualization}. Furthermore, because humans have a very limited ability to recall prior values when serially inspecting data, this internal representation suffers from inaccuracies caused by forgetting or misremembering past data. 
As Munzner stated, ``Vis allows people to offload internal cognition and memory usage to the perceptual system''~\cite{munzner2014visualization}. It does so by creating an external representation of the data that humans can comprehend more easily.

\oursystem{} aims to create a debugging system that shifts the perspective from debugging programs through several serial views to take the previously described visualization first perspective on debugging. It focuses on giving programmers an overview of the data within their program first and then providing them tools that allow them to delve into the details as desired.  The ``Task Analysis'' section provides a more in depth inspection of the goals of \oursystem{} and the tasks necessary to support those goals. 

% \oursystem{} aims to take a visualization first approach to debugging.  

% \rf{Some description of why we are making this system, then reference task analysis for specific goals and tasks of the system - tie in to that one reviewers questions about global views?}

\paragraph{Audience}
The audience dimension defines who will use the visualization.
\oursystem{} aims to help python programmers understand their programs and diagnose misbehavior's in the programs they are running. While our prototype currently supports the visualization of program traces of a moderate size (around 225,000 recorded function calls and variable assignments), we believe that the design of Anteater is appropriate for general programming tasks in Python. 

% \rf{Description of the intended audience and scope of programs we intend to support. Highlight, that it is a prototype that is not intended to be used for large scale software systems.}

\paragraph{Target}
The target dimensions defines what aspects of the program are visualized. \oursystem{} creates a trace as the program executes. \oursystem{} focuses on collecting internal program values, such as variables and expressions, throughout the entire execution of the program. Additionally, these traces capture the calling and looping structure of the execution. The details of the tracing infrastructure of \oursystem{} are discussed in the section ``Tracing Infrastructure and Data Organization''.

% \rf{Target is program traces and internal values. Calling \& looping structure. Values throughout the entire execution}

\paragraph{Representation}
This dimension defines how to convey the target information to the user. \oursystem{} leverages well understood visualizations of each type of data collected to present the data to the programmer in an easily understandable way. It then pairs these visualizations with interactions that allow people to filter down to areas of interest in their program values and view details as desired. The visual design is discussed in depth in the section ``Anteater's Visualization Design''.

\paragraph{Medium}
The medium dimension defines where this information is displayed. We intend \oursystem{} to be displayed in color on a laptop screen or an external monitor.

\section{Task Analysis}
\label{sec:tasks}

In this section, we discuss \oursystem{}'s goals. The original inspiration for our goals came from Omnicode~\cite{kang2017omnicode} and the Coz profiler~\cite{curtsinger2015coz}.  We further refined our goals after exploring additional related work, characterizing the problem with Munzer's framework, and reflecting on our own experiences with respect to program debugging and understanding. The final goals below were derived after several iterations of system design and goal refinement. 

\paragraph{{\color{myOrange}G1: Identifying the source of unexpected execution behavior}} When programmers write and execute programs, they have some expectation of how their program should be behaving, e.g. what functions should be called and when. As a result, one goal of debugging is to identify what is causing an execution to deviate from what the programmer expected. To support this goal, debugging tools need to provide a view of the execution structure (see Features column of Table~\ref{tab:comparisons}). Furthermore, this goal encompasses the subset of the \textbf{search} actions identified in the previous section that correspond to understanding the execution structure of a program. For example, a programmer may want to \textit{lookup} a specific function call, \textit{locate} an erroneous function call, \textit{browse} a specific area of the execution structure, or \textit{explore} the overall structure of the execution.  

\paragraph{{\color{myBlue}G2: Identifying the source of unexpected values and trends}} Similar to {\color{myOrange}G1}, programmers typically have a general ideas about what variable values they should observe during the execution of a program and thus desire to identify the root cause of unexpected values in the execution. To help programmers identify patterns and trends in the values of a variable, tools need to provide views of variables over the entire execution of the program.  This corresponds to the "Single Variable, Whole Time" column in Table~\ref{tab:comparisons}. In addition, keeping these values in context of the execution structure, allows programmers to isolate areas of interest in the execution.  Whereas {\color{myOrange}G1} encompasses the subset of the \textbf{search} actions corresponding to understanding the execution structure of a program, this goal encompasses those corresponding to understanding the internal variable values of a program. For example, a programmer may want to \textit{lookup} a specific instance of a variable, \textit{locate} an erroneous variable calculation, \textit{browse} instances of variables at a particular point in the execution, or \textit{explore} the overall trends of a variable throughout the execution.

\paragraph{{\color{myGreen}G3: Explore the behavior of an unfamiliar and/or complex piece of code}} This goal encompasses a wide range of exploratory debugging and understanding tasks. We designed it to be general enough cover any programming situation that did not fit into the first two goals. For example, programmers are often tasked with understanding code written by someone else.  Typically, this is no easy task and requires a significant amount of effort on the part of the programmer. Viewing the structure of the execution along with trends of variables throughout the entire execution serves as a starting point for understanding the behavior unfamiliar code. Similarly, programmers use well known but complex analysis algorithms that they write but do not fully understand how the algorithm operates. Understanding these algorithms is a difficult task that requires effort similar to understanding code written by someone else. This goal aims to encompass programming tasks like these.  All debugging and understanding tools attempt to support this goal and as a result, all views and features described in Table~\ref{tab:comparisons} support this goal.  This goal encompasses the subset of the \textbf{search} actions corresponding to understanding the general behavior of a program. This goal often corresponds to the \textit{explore} and \textit{browse} actions where the target is not concretely defined.

\paragraph{}Under the framework of Lam et al.\ \cite{lam2018bridging},  {\color{myGreen}G3} falls into the ``Discover Observation'' category and {\color{myOrange}G1} and {\color{myBlue}G2}, fall into the ``Identify Main Cause'' category. From these goals, we derived several sub-tasks required to support the goals.

\paragraph{T1: Inspect all instances of a variable or expression} It is often useful to look at all of the values that a variable or expression takes on to determine if it is behaving as-expected and to identify any erroneous values (supporting {\color{myBlue}G2}). Additionally, in an unfamiliar or complex program, it helps create a general understanding of the variables behavior (supporting {\color{myGreen}G3}). This task corresponds to the low-level actions \textit{summarize} (e.g. view the trends of a variable) and \textit{identify} (e.g. inspect an erroneous value) as described in the previous section. 
% \rf{Most tools support this in some form. With print debugging, simply printing a variable whenever it is assigned to fulfills this task. Similarly, stepping through an execution and inspecting a variable at every step technically fulfills this task, although it does not allow the user to view all the values at once.}

\paragraph{T2: Identify what functions are called at runtime} Often it is not clear from the static source code which functions will execute and when.  However, identifying which functions are actually called during an execution is crucial for understanding how a program is operating (supporting  {\color{myGreen}G3}) and identifying unexpected execution behaviors (supporting {\color{myOrange}G1}). Providing an overview of the execution (corresponding to the \textit{summarize} action) allows people to see which functions are called at runtime and allows them to isolate misbehavior (corresponding to the \textit{identify} action).

% This task corresponds to the \textit{summarize} (e.g. view what function) and \textit{identify} (e.g. inspect an erroneous value) as described in the previous section. 
% \rf{In Table~\ref{tab:comparisons}, this refers to the "Execution Structure Visualization" column. With manual effort, this could also be supported with print debugging by printing a note when each function is called and with breakpoint debugging by stepping through the entire execution. } 

\paragraph{T3: Identify dependencies for a variable} Understanding dependencies is crucial when trying to understand the behavior of a program.  Identifying how a value is calculated, including the execution path required to complete the variable's calculation, allows programmers to better understand the underlying nature of the value in question (supporting  {\color{myGreen}G3}).  Such insight can lead to finding the cause of an unexpected value (supporting  {\color{myBlue}G2}).  This task supports the \textit{identify} action in relation to viewing the dependencies of a specific instance of a variable.
% \rf{TODO: exisitng work?}

\paragraph{T4: Identify interesting subsets of values} Given a variable or expression, it is important to be able to identify the subset of values that correspond to interesting behavior. For example, if certain values indicate a failure in the program, they need to be identified so the surrounding values can be examined to understand the cause of the behavior.  This task supports {\color{myBlue}G2} and {\color{myGreen}G3} as well as the \textit{identify} action. 

\paragraph{T5: Observe relationships between values} When debugging a program, programmers often investigate relationships between variables (supporting the \textit{compare} action).  For example, if variable $x$ changes, how does variable $y$ change? While these relationships may not be explicitly defined by the code, i.e., $y$ may not directly depend on $x$, they often provide meaningful information to the programmer. Uncovering such relationships contributes to program understanding (supporting {\color{myGreen} G3}). 

\paragraph{T6: Maintain context between runtime state and static source}  When trying to debug and understand a program, maintaining context with the actual code is critical.  If the programmer is manually instrumenting print statements, they also must codify contextual information to derive insight, e.g., representing the location of a variable's modification. This task supports {\color{myOrange}G1}, {\color{myBlue}G2}, and {\color{myGreen}G3}.  

\paragraph{}A system that supports all of these tasks needs to track the execution structure of the program along with variable and expression values in the context of its execution.  An execution trace fits this need as it naturally tracks the execution structure of a program and can be modified to also collect values.  Once a system collects this data, it must present it in a way that allows for easy navigation through the data while supporting the defined tasks. We argue that visualization best way presents this information because it is known for providing overviews and context, highlighting relationships, and facilitating the filtering down to subsets of interesting information, all of which are needed to support these tasks.  \oursystem{} takes a visualization approach to program debugging and understanding that satisfies these goals through execution traces and visualizations.  Currently, \oursystem{} deals solely with single-threaded programs but we expect that this task analysis would need to be extended to satisfy our goals for multi-threaded programs.

\section{Tracing Infrastructure and Data Organization}
To support the goals and tasks defined in above, an execution trace with accompanying variable and expression values must be collected.  \oursystem{} implements a tracer that automatically instruments source code to collect its execution trace. Implemented in Python, the tracer relies solely on the Abstract Syntax Trees (AST) to facilitate the transformation of the source code.  While \oursystem{} currently only works with Python programs, the same principles can be implemented in any language that has the ability to transform source code in a similar way. After transforming the source code, \oursystem{} runs the program, generates the trace file, and organizes the data in a way that allows for easy creation of interactive visualizations. Fig.\ \ref{fig:sysOVerview} illustrates how the system operates. 

\subsection{Tracing Programs}

This section goes into depth on part (a) and (b) of Fig.~\ref{fig:sysOVerview}. First, it discusses how people can specify traces through the \oursystem{} front-end. Then, it discusses how \oursystem{} turns this trace specification in to program trace. 

\paragraph{Specifying a Program Trace} 
To fulfill T1 (inspect all instances of a variable or expression), \oursystem{} allows programmers to define which variables and expressions to track, through interactions with the source code. Additionally, to eliminate unimportant functions form the trace, people may specify functions and libraries to exclude from the trace. Together, these two pieces create a trace specification. This corresponds to part (a) of Fig~\ref{fig:sysOVerview}.
%Note, we differentiate between variables (e.g. anything on the left of an ``=" sign, parameter definitions, etc.) and expressions (e.g. anything on the right of an ``=", function calls, etc.) to ease tracing efforts in the back-end.  
\oursystem{} also allows people to define additional custom expressions associated with their chosen variables that it evaluates and records each time it records the corresponding variable.

\oursystem{} best supports numerical values but has limited support for strings and boolean values.  While it cannot directly visualize lists and matrices, information about either structure can be tracked using custom expressions (see section "How to Handle Objects"). Once the programmers complete the trace specification, \oursystem{} passes it to the tracer in the backend for processing.

 Note, the tracer will only collect the variables and expressions defined in the trace specification.  We explicitly chose to do this because collecting the entirety of data associated with every variable in the program leads to the collection of massive traces filled with a significant amount of irrelevant/unnecessary data. Many variables residing in code have little importance in describing the program's behavior. Thus, \oursystem{} allows the user to select the important variables to track. This decision discussed more in the Discussion section. 
 
% Similarly, collecting all function calls leads to a large collection of unimportant information. To help reduce clutter from unimportant function calls, we add a predefined list of functions and libraries to ignore (e.g., \texttt{math}, \texttt{numpy}, \texttt{print}, \texttt{len}) and allow users to add functions of their own to exclude from the trace.  This allows users to reduce clutter and remove uninteresting/unimportant structures from the trace to better highlight the important structures.

\begin{figure*}
    \centering
    \includegraphics[width=\linewidth]{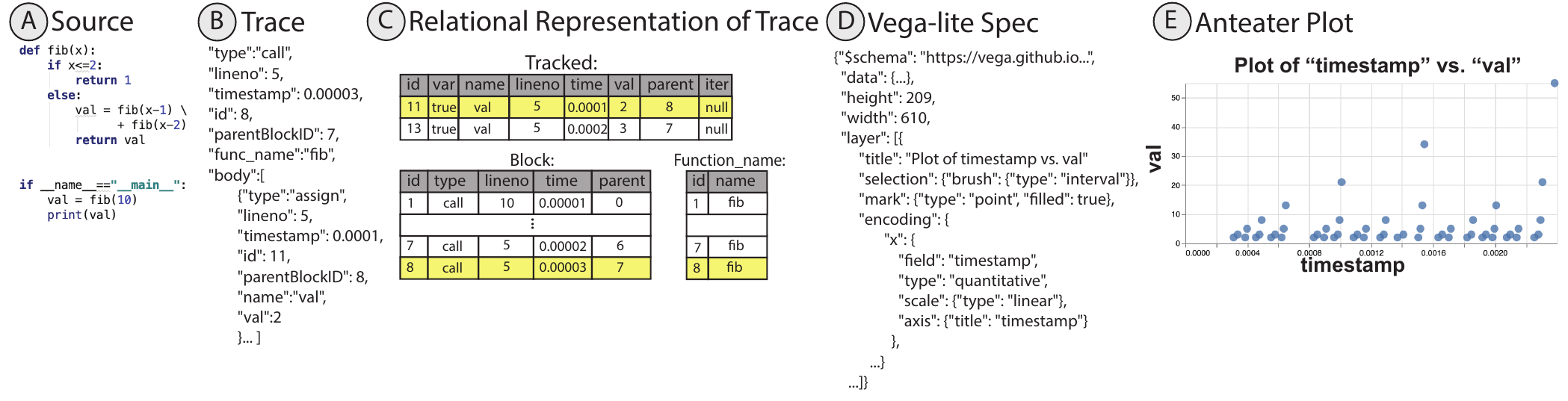}
    \caption{An overview of how \oursystem{} goes from source code to visualization. (A) shows the initial source code. We are going to track the variable ``val'' After instrumenting the source code, as demonstrated in Fig.~ \ref{fig:sysOVerview}.  The instrumented program creates a trace cell as shown in (B).  Anteater then puts the JSON into a SQL table as shown in (C).  From there, \oursystem{} queries the table to select all points from ``Tracked'' that have the name ``val'' and passes them to \oursystem{}'s Vega-lite generator which generates a Vega-lite specification (as shown in (D)) for the corresponding plot.  \oursystem{} then renders the specification to create a scatterplot of those points over time (shown in (E)). }
    \label{fig:translate}
\end{figure*} 

 \paragraph{Anteater's Tracer} When a user chooses to create a trace, the \oursystem{} back-end is passed a trace specification containing a list of variables and expressions to track and a list of functions and libraries to exclude from the trace.  The tracer indexes through these lists and determines the scope in which each item resides to ensure that it only tracks/excludes the specified items.  For example, if two disjoint functions both define variable $x$, the tracer will only track the one the user selected.

Once \oursystem{} determines the scope of each item, the tracer uses the Python \textit{ast} library to parse the source code into its AST. It then performs a series of traversals of the AST to collect information about the source code and transform the program to trace the execution and desired values. 

In the first traversal through the AST, no transformations occur. Rather, \oursystem{} collects information about functions, loops, and dependencies.  For functions and loops, it collects the lines at which the function definition or loop begins and ends. This information enables more detailed linking between visualizations and source code. For dependencies, the tracer traverses through the code and, for each variable, stores functions and variables on which it directly depends in the source text. 

Once all of the static data has been retrieved from the source code, \oursystem{} begins transforming it.  A second traversal through the AST transforms the code to isolate all function calls from their respective expression statements and expand list comprehensions into \texttt{for} loops. \oursystem{} pulls all function calls that do not stand alone out of their expressions and assigns them to a temporary variable that replaces the  call in the original expression (e.g., $x = 2* f()$ becomes $tempF = f()$; $x = 2*tempF$).  This allows \oursystem{} to easily capture when and in what order functions are called.  

Next, the tracer performs the main transformations to insert the instrumentation that collects the trace.  As the tracer traverses the AST, it always pauses at assignment, call, and loop nodes. When it reaches an assignment node, it checks the trace specification to determine if the target variable needs to be tracked.  If so, it inserts new nodes into the AST that record the value of the variable after assignment.  

When the tracer reaches a call node, it first checks if the trace specification excludes the function.  If not, the tracer wraps the call with AST nodes to record the entry into and exit from the call.  
% Because we separate function calls into their own statements, a function call statement fully executes before the next function call starts.  This allows all bookkeeping for a call to be completed before the next call executes. 
A simplified example of this transformation is shown in Fig.\ \ref{fig:sysOVerview} (c).

When the tracer reaches a loop, it creates a counter to track the iteration of that loop and inserts new instrumentation to record the start of the loop. As it traverses the body of the loop, any time the tracer creates a new record, it records the iteration in which that record occurred.  Tracking the iteration binds together groups of records in the trace that occurred in the same part of the execution (i.e. records that occurred in the same iteration).

% \begin{figure}
%     \centering
%     \includegraphics[width=\linewidth]{translate_v3.pdf}
%     \caption{An overview of how \oursystem{} goes from source code to visualization. (A) shows the initial source code. We are going to track the variable ``val'' After instrumenting the source code, as demonstrated in Fig.\ \ref{fig:sysOVerview}.  The instrumented program creates a trace cell as shown in (B).  Anteater then puts the JSON into a SQL table as shown in (C).  From there, \oursystem{} queries the table to select all points from ``Tracked'' that have the name ``val'' and passes them to \oursystem{}'s Vega-lite generator which generates a Vega-lite specification (as shown in (D)) for the corresponding plot.  \oursystem{} then renders the specification to create a scatterplot of those points over time (shown in (E)). }
%     \label{fig:translate}
% \end{figure} 

Lastly, the tracer transforms the program to record expressions.  Unlike variables, expressions occur in a variety of AST nodes.  As the tracer visits each node, it checks if the line containing the node also contains a tracked expression. If it does, the tracer isolates the expression from the line, assigns it to a temporary variable, and then replaces the expression in the original line with the temporary variable. This ensures that the expression only executes once and that the trace records its exact behavior during the execution of the program.  

Once \oursystem{} completes the instrumentation, it compiles the AST into an executable program, which generates the trace as it executes. 

\subsection{Data Organization}
Fig~\ref{fig:translate} illustrates how we go from the source code to visualizations. After \oursystem{} instruments the source code, it runs the modified program and creates the trace file.  \oursystem{} writes the raw trace as a simple JSON file, shown in Fig.~\ref{fig:translate} (B).  This allows it to easily capture the hierarchical structure of the execution as well as record data about program blocks as attributes in the corresponding JSON block. \oursystem{} then passes the trace to the front-end. While convenient for collecting the trace, JSON is less convenient and flexible for querying the trace which limits the range of possible visualizations and interactions. 
%Extracting instances of variables from the JSON structure requires traversing the entire structure. This makes even the most basic queries, simply selecting data from the structure, inefficient. Furthermore, more complex interactions such as filtering a variable or joining two variables, result in complicated traversals of the JSON structure that further decrease the efficiency of queries.  
To support more complex visualizations and interactions, \oursystem{} converts the JSON trace into a SQL database.    

As shown on the right side of Fig.~\ref{fig:sysOVerview} (d), \oursystem{} converts the JSON trace into SQL tables.

The primary two tables store (1) the attributes of the nodes in the execution tree (the ``block'' table), and (2) the attributes of all instances of tracked variables and values (the ``tracked'' table). 
% the primary two being ``block'' and ``tracked''.  The ``block'' table stores information about all execution blocks that do not correspond to tracked values, such as function calls and loop iterations.  It includes columns storing the id, type, line number, timestamp, and parent id of the block.
Fig~\ref{fig:translate} B-C demonstrates how to convert form JSON into the corresponding SQL tables.  

% \oursystem{} represents each of these blocks as a bar in the generalized context tree (GCT) plot, shown in Fig.~\ref{fig:fibOverview}. To create the GCT, \oursystem{} must recreate the hierarchical structure of the trace.  It does this by, starting at the root block that represents the whole module, querying the database for all child blocks (found via the parentID attribute), adding them as children to the current block, and repeating for each child block. In essence, this re-builds the tree in a manner similar to depth first search. 

% Using the parent id of each block, we build the hierarchical structure used to create the generalized context tree by starting at the root, querying the SQL database to find all of the blocks that have the root as their parent, adding each child node to the hierarchy, visiting each child node and repeating this process.

% The other primary table, the ``tracked'' table, stores the occurrences of the variables and expressions the user tracked. This table stores the id, name, line number, timestamp, value, parent id, iteration number, and a boolean value indicating if it is a variable (as opposed to an expression). \oursystem{} queries this table to build the visualizations. For basic, unfiltered visualizations, \oursystem{} simply queries this table for all records with a certain name.  For filtered queries, it either specifies a range or a specific set of id's that the record can take on using the \texttt{WHERE} clause. 

Additional tables exist, such as ``function\_name'' and ``for\_loop'' that store additional information about certain types of blocks.  The ``custom'' table stores the values of custom expressions that are collected alongside the variables and expressions selected in the source code.  

Converting the trace to SQL yields several advantages.  First, querying becomes much simpler.  For basic visualizations, we now must simply write a \texttt{SELECT} statement to gather all instances of a tracked variable.  To filter instances, we can simply add a \texttt{WHERE} clause to the SQL statement. Similarly, joining two variables becomes much simpler through the use of \texttt{JOIN}. Table\ \ref{fig:data_types_vs_vis} shows a table of visualizations supported by \oursystem{} and the corresponding SQL query keywords used to collect the data. 

Second, \oursystem{} supports any visualization for which there exists a SQL query to select the appropriate data.  In other words, forming the proper query becomes the only restriction to the range of possible visualizations. While the current implementation only supports a few visualizations, we could easily extend it to support others.

\begin{table}[htbp]
    \centering
    \small
\begin{tabular}{|c|c|c|}
\hline
    Data Type & Plot Type & Query\\
    \hline
    Q & Histogram & \texttt{SELECT}  \\
     \hline
      N &  Bar plot & \texttt{SELECT} \\
     \hline
     QxQ & Scatter & \texttt{SELECT}, \texttt{JOIN}\\
       \hline
     QxQxQ... & Parallel Coordinates & \texttt{SELECT}, \texttt{JOIN}\\
     \hline
     N, Q, QxQ & Small Multiples & \texttt{SELECT}, \texttt{JOIN}, \texttt{SORT ON}\\
     \hline
\end{tabular}
    \caption{The above table shows the current visualizations supported and the SQL queries used to create these visualizations.  We use "Q" for quantitative data and "N" for nominal.}
    \label{fig:data_types_vs_vis}
\end{table}

The last advantage comes from the decoupling of the visualizations and the data representation.  The specification of the visualizations does not inherently depend on the representation of the data. A SQL query simply returns a list of datapoints for \oursystem{} to use in the visualization. Because of this, we easily adapted \oursystem{} to use Vega-Lite~\cite{satyanarayan2016vega} specifications to generate visualizations. Furthermore, new visualization implementations can be plugged in with minimal effort to adapt them to fit into \oursystem{}. This further increases the extensibility and flexibility of \oursystem{}. 

\subsection{Generating Vega-lite Specifications}
As mentioned previously, a SQL query simply returns a list of datapoints.  \oursystem{} then simply needs to generate a Vega-lite specification appropriate for the specified data (the final step of Fig.~\ref{fig:sysOVerview}(d)). 
A snippet of a generated specification is shown in Fig.~\ref{fig:translate} (D) with the corresponding plot in (E).  Leveraging the power of Vega-lite allowed us to easily create clean, interactive visualizations that are customized to best present the data selected by the programmer.

\section{\oursystem{}'s Visualization Design} 
\label{sec:visDesign}

\oursystem{} presents a new way of exploring and interacting with program executions helping users to gain a deeper understanding of the inner-workings of their programs that they cannot get from traditional tools.  In the previous section, we discussed how \oursystem{} creates the execution trace. Here, we describe the visualization design of \oursystem{} and the features that facilitate the exploration of the execution trace. As we walk through the design, we will describe the features in context of a simple Python program that runs a recursive Fibonacci function. In addition, we use Yi et al.'s categories of interactions~\cite{DBLP:journals/tvcg/YiKSJ07} to classify our interactions and further validate our design.  \oursystem{} uses Vega-lite to generate all visualizations, with the exception of the generalized context tree.

\begin{figure}[t]
    \centering
    \includegraphics[width=\linewidth]{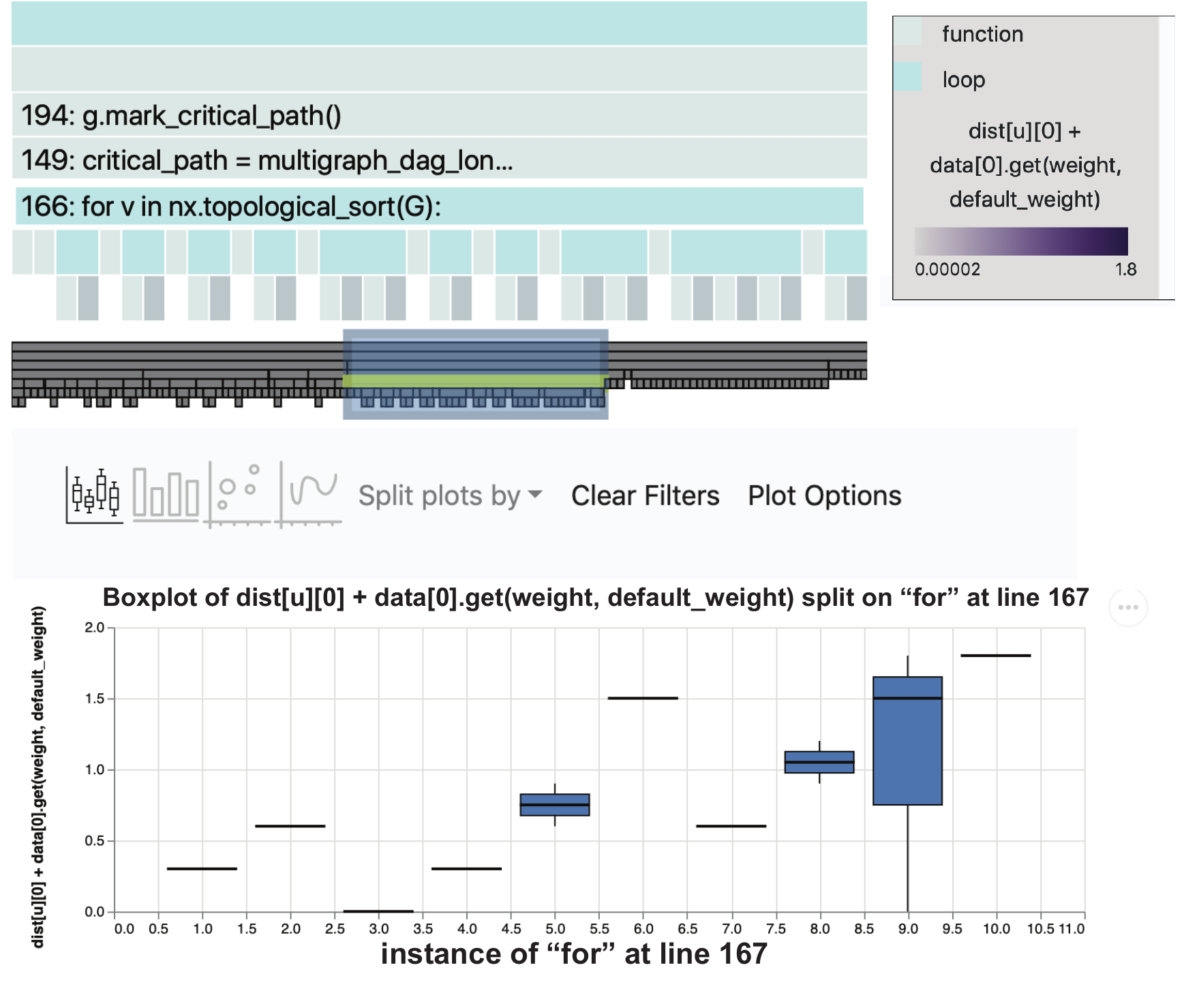}
    \caption{An example of \oursystem{} splitting the data by a structural element. \oursystem{} splits the data by instances of a for loop at line 167, which corresponds to iterations of the loop at line 166 (the selected block in the generalized context tree). The plot shows one boxplot per loop instance.}
    \label{fig:grouping}
\end{figure}

\subsection{Visualizing Program Data}
Once the tracer returns the execution trace, \oursystem{} generates interactive visualizations.  Two types of visualizations are provided:  a view of the execution structure, which we call the generalized context tree, and a visualization of the variable values.  For ease of use, Anteater provides well understood visualizations of the program information but can be easily extended to support more complex/custom visualizations. 

\subsubsection{Generalized Context Tree}
The generalized context tree (GCT), shown on the right side of Fig.\ \ref{fig:fibOverview}-A and in Fig.\ \ref{fig:fibOverview}-B, provides an overview of the execution structure. The visualization has its origins in flame graphs and icicle plots.  We chose this type of visualization because it is well known  and understood for visualizing traces.  In our setting, each rectangular block in the plot represents one of three things: a function call, a loop, or a variable assignment. The icicle plot shows the hierarchy so that, for a given block, everything that is within that blocks bounds below it, is a child which means it executed within the code of the parent block (i.e. in that call or loop iteration).  For example, in Fig.\ \ref{fig:fibOverview}-A, the block in the second row labeled ``10: val = $\dots$'' is the initial call into the Fibonacci function and everything below that happens within that call.  The generalized context tree can be used to determine which functions executed and when, fulfilling T2 (identify which functions are called at runtime).  

As we move from left to right in the plot, we are increasing in time; everything to the left of a block was fully executed before that block.  This allows users to easily read the visualization and understand when blocks are executed relative to other blocks.

The GCT highlights a single variable corresponding to the variable on the x-axis of the plot.  When the user assigns a variable to the x-axis, the GCT colors all blocks in the tree corresponding to that variable (which reside at the leaf level) by the value of the corresponding instance. Positive values range from white (low) to purple (high), while negative values range from white (least negative) to orange (most negative). In Fig.\ \ref{fig:fibOverview}-A, \oursystem{} colors the leaf nodes representing the variable "val" with varying shades of purple.  Deeper leaves are shaded much lighter, which indicates small values at those instances; this corresponds to the deepest Fibonacci calls returning the smallest values.  Coloring blocks in this way shows the behavior of values in the context of the whole execution.  Every other variable or expression that appears in the trace still appears in the generalized context tree but \oursystem{} colors them gray to keep focus on the selected variable. 

Before creating the GCT, \oursystem{} must organize the data into a hierarchy that it then passes to the D3 library to generate the visualization. To organize the data into the hierarchy, \oursystem{} starts at the root block that represents the whole module and queries the database for all of its child blocks. It then adds these blocks as its children to the hierarchical data structure and repeats this process for each child block. In essence, this re-builds the tree in a manner similar to depth first search.

\begin{figure*}[t]
    \centering
    \includegraphics[width=\linewidth]{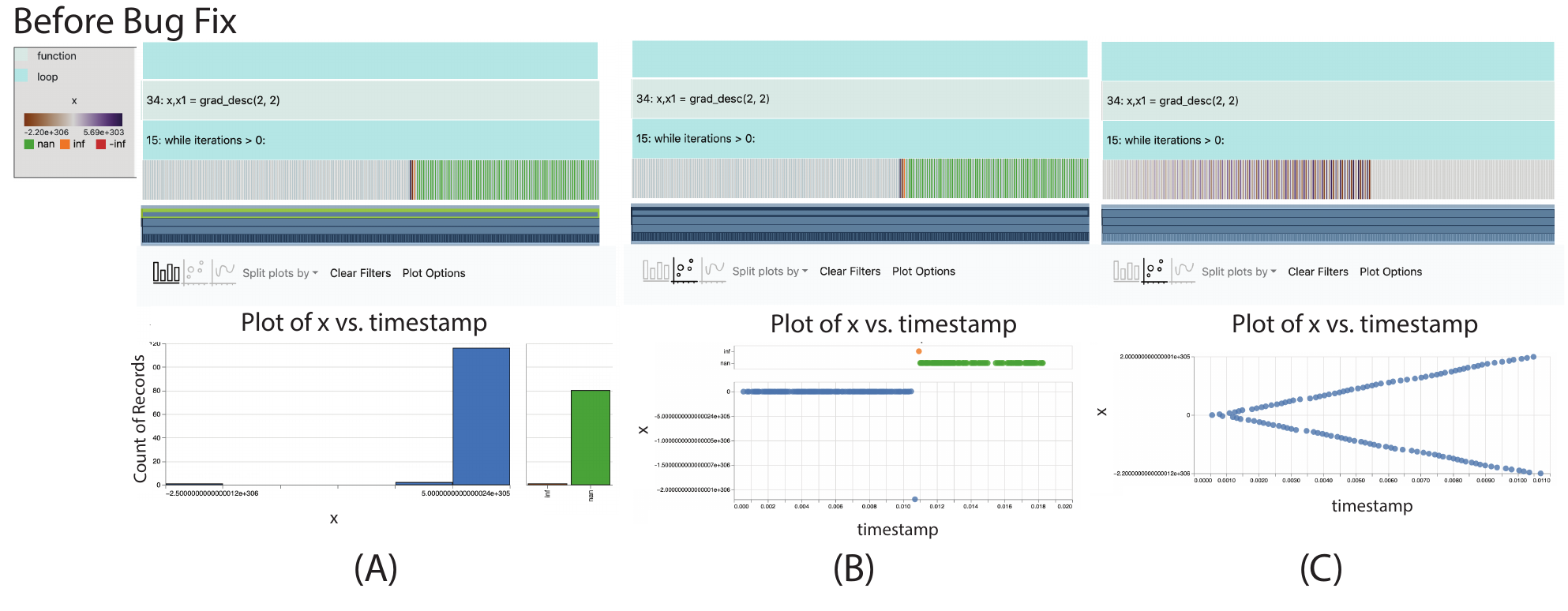}
    \caption{Debugging Gradient Descent with \oursystem{}. In (A) it is immediately apparent in both the generalized context tree and the histogram that there is a bug causing NaN's, shown in green in both the histogram and GCT (NaN means ``Not A Number'', special floating-point values that indicate numerical failures). In (B), we switch to the scatterplot view to see how the values behave before they become NaN. The values are mostly centered around zero before becoming an extremely small negative, then going to infinity and becoming NaN. We suspect that the values centered around zero are not actually zeros so we filter the values in the scatterplot to allow us to zoom in on them and switch to a symmetric log scale, shown in (C). Now we see that the values are oscillating which suggests the problem of exploding gradients caused by a training rate that is too large. Fig.~\ref{fig:gdOverview2} shows the \oursystem{} visualizations after correcting the bug.}
    \label{fig:gdOverview}
\end{figure*}

\begin{figure}
    \centering
    \includegraphics[width=.9\linewidth]{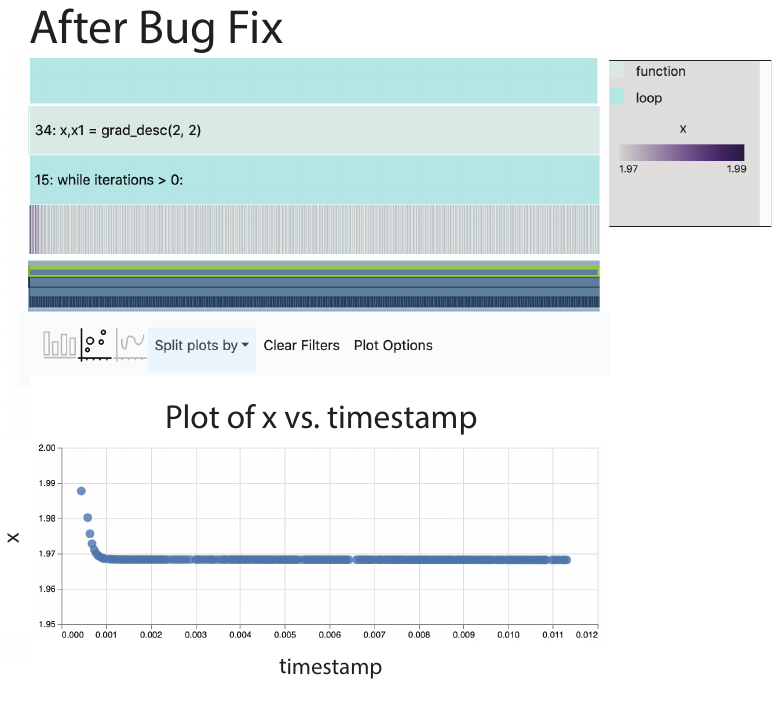}
    \caption{Debugging Gradient Descent with \oursystem{}. The plot and generalized context tree after we correct the bug from Fig.~\ref{fig:gdOverview}. To correct the bug, we reduce the training rate and can see that the value quickly converges as expected.}
    \label{fig:gdOverview2}
\end{figure}

\subsubsection{Variable Value Plots}
The second visualization provided by \oursystem{}, is a plot of tracked variables. Similar to when creating the trace specification, programmers add tracked variables and expressions to the plot by right clicking and selecting to add it.  \oursystem{} queries the database to retrieve the specified variables. When \oursystem{} initially reads in the trace, \oursystem{} checks each tracked variable and expression to determine its type (quantitative or nominal). Thus, when creating a plot, \oursystem{} first checks the data types of each involved variable before looking up the plot type appropriate for the selected variable(s) (based on Table~\ref{fig:data_types_vs_vis}).  Once \oursystem{} determines the correct plot type, it begins generating the Vega-lite specification.  Initially, it creates the base layer that sets the mark for the plot (bar, point, line, etc.) and plots the initial data. In this layer, \oursystem{} performs any necessary filtering and transformations (e.g. aggregation for histograms and filtering out non-numeric values in quantitative data such as "NaN" values).  If quantitative variables have non-numeric values \oursystem{} will concatenate additional subplots (horizontally or vertically depending on which variable contains the values) to show these values.  Vega-lite allows \oursystem{} to sync the axes of the subplots with the main plot in the base layer, as in Fig.~\ref{fig:gdOverview} A and B. This builds the base visualization for the specified variables.

\subsection{Interacting with the Trace Visualizations}
\oursystem{}'s interactions are key in helping users get a better understanding of their program.  We organize our interactions based on Yi et al.'s categories of interaction:
Select, Explore, Reconfigure, Encode, Abstract/Elaborate, Filter, and Connect.

\paragraph{Select and Connect} \oursystem{} provides  interactions that connect related views in the following way: interactions to link the generalized context tree and the plot view (in both directions) and interactions to link the visualizations to the source code. Additionally, the interactions linking the generalized context tree to the plot view also serve to select portions of the execution data that are of interest.   

\oursystem{} provides interactions on the plots and the GCT to link the two together.  When a user selects a block in the GCT, the values shown in the plot filter down to include all values in the subtree rooted at the selected block.  In addition, to provide global context, the plot shows the values from the subtree rooted at the parent of the selected block. As shown in the histogram in Fig.\ \ref{fig:MPI}-B, \oursystem{} colors the bar representing the selected instance(s) blue while the coloring rest of the bars gray for context. In the scatterplot, it colors the points representing selected instances while leaving the rest gray.  \oursystem{} also provides  linking from the plot back to the GCT. In the histogram, selecting a bar highlights the corresponding blocks in the tree, as shown in Fig.\ \ref{fig:MPI}-A.  In the scatterplot, brushing over a set of points highlights the corresponding blocks in the trees, as shown in Fig.\ \ref{fig:fibOverview}-B where the red blocks in the tree correspond to the brushed points. \oursystem{} enables these selections by adding specific parameters to the Vega-lite specification. These parameters specify the type of selections available (e.g. brushing or clicking) and the visual effects of the selections (e.g. changing color or opacity of unselected points). Furthermore, Vega-lite's data listeners allow \oursystem{} to monitor these selections and update linked views accordingly.  These interactions support T4 - identify interesting subsets of values - by allowing the user to pinpoint interesting values in the plots and locate them in the execution. 

Additionally, when exploring the execution, it is important to connect back to the source code to maintain the context of the execution.  On its own, the generalized context tree is fairly abstract.  To provide necessary context, when the user selects a block in the generalized context tree, the source code jumps to, and highlights, the corresponding section of the code. If it corresponds to a function call whose definition resides in the source file, it also highlights the corresponding function.  This interaction, paired with a preview of the corresponding source code on the blocks, supports T6 - maintain context between runtime state and static source - by allowing users to navigate the execution trace without forgetting their place in the source code.

\paragraph{Explore}
 \oursystem{} supports two ``explore'' interactions: faceting values into groups and inspecting dependencies.  

The first interaction, faceting values into groups, enables people to view distinct subsets of a variable. \oursystem{} provides grouping capabilities that allow the user to facet the data into groups and create either a series of box and whisker plots on the same axes (one for each group) or small multiples of plots.  The data can be split on either a related variable/expression from the trace (such as a boolean value) or a repeated structure in the execution, such as a loop, where each instance of the structure contains multiple instances of the tracked variables/expressions. For example, in Fig.~\ref{fig:grouping}, \oursystem{} splits the plot on the outer loop and creates a box and whisker plot for each instance of the inner loop. 

The second ``explore'' interaction supports the inspection of dependencies. To support T3 (identify dependencies for a variable), \oursystem{} determines what dependencies could exist for any instance of a variable. To find all dependencies for a variable, \oursystem{} accesses the variables dependency list generated during tracing, and then, for each dependency in that list, it accesses their dependency lists.  This continues until \oursystem{} builds a comprehensive list of all possible dependencies. 

After creating the list of dependencies, \oursystem{} uses context from the execution trace to eliminate some possibilities and present the remainder to the user. When a user selects a block in the generalized context tree that represents a variable, \oursystem{} checks 2 sets of blocks: (1) any siblings of the selected block that were fully executed before it and (2) the siblings of all ancestor blocks of the selected block that were fully executed prior to the selected block. From these sets of block, \oursystem{} finds any blocks that are on the list of possible dependencies.  For any block that is on the list, it is highlighted in the generalized context tree to show the user the user on which parts of the context tree that selected block depends. 
% When a user selects a block representing a variable in the generalized context tree, \oursystem{} checks if the prior siblings as well as the siblings of any ancestors of that block are in the list of possible dependencies.  If they are, their blocks are highlighted in red to show the user on which parts of the context tree that block depends. 
This allows the user to quickly get an idea of which entities may contribute to that specific instance.  In Fig.\ \ref{fig:fibOverview}-B, the selected instance of ``val'' depends on the prior two calls to ``fib''.

\paragraph{Reconfigure}: \oursystem{} supports reconfiguration by allowing users to add multiple variables to a plot (supporting T5 - observe relationships between values). If the variables are compatible, \oursystem{} plots them against each other in either a scatterplot or parallel coordinates (depending on the number of variables), allowing the user to observe their relationship. Compatible variables share a common ancestor and have 1-1 instances within that ancestor. \oursystem{} provides an options menu that allows programmers to swap or change the scales on axes using the "Plot Options" menu.

 \paragraph{Encode} Depending on the type of data presented, \oursystem{} allows people to encode the data in a multiple ways.  People can click on the icons above the plot to switch between the different plot types available for that datatype. Additionally, \oursystem{} gives them controls to rearrange the axes of the plots as well as change the scales. 

\paragraph{Filter}
\oursystem{} supports three types of filter interactions on the plot and the generalized context tree to help people filter out unimportant information and emphasize important parts of the execution, which helps support T4 (identify interesting subsets of values). The first type of filtering was mentioned above where clicking on deeper nodes in the context tree filters the value plots. Through this interaction, users can filter down the plot to interesting subsets of the data.   

In the scatterplot, users can brush over a subset of points, right click, and select to filter out the values not in their brush.  \oursystem{} then removes all other points from the plot, effectively zooming in on selected points, and grays out any block not on the path to a shown point. Examples of this can be seen in Fig.\ \ref{fig:gdOverview}-C and Fig.\ \ref{fig:MPI}-C. Similarly, in a bar plot or histogram, users can select bars and filter down to the corresponding values in the same manner.

One last way users can filter the visualization is by hiding parts of the generalized context tree.  Right clicking on a block in the tree will expand the block to take up the entire width of the interface, increasing the size of all of its children and thus making them easier to see.  However, in doing this, users might lose context of where they are exploring with respect to the execution.  To retain this context, we add a smaller, grayscale version of the generalized context tree with a highlighter bar over it. When the user zooms in on a block, the highlighter narrows to indicate its place in the overall context tree.  It also highlights the selected block in yellow, as well as any other blocks that are highlighted in the generalized context tree (from dependencies and brushed values). This allows users to see highlighted blocks even if they are outside of the visible portion of the generalized context tree. In Fig. \ref{fig:grouping}, we zoomed in on the loop at line 166, but we see our location with respect to the whole generalized context tree in the context bar.

\subsection{How to Handle Objects}
While \oursystem{} will not directly collect objects, it provides a way for users to collect the information that interests them from the object. To do this, the user locates the place in the program where they wish to inspect the object. At this point, they choose to create a custom expression for Anteater to record that accesses the data in the object that interests them.  Each time the execution reaches this point, Anteater will evaluate and record the value of the expression.  This enables users to indirectly gather all of the information from objects that they wish to inspect without directly collecting the entire object.

The central challenges with collecting entire objects are the detection of every modification to the object and visualizing all information within an object. The first challenge would require Anteater to detect every time the object is mutated and record the new state of the object.  Not only is the detection a difficult task, but the collection of all mutations of the object will inevitably lead to unmanageably large trace files.  The second challenge would require additional input from the user on how to design the visualization of the object given the information it contains.   Rather than have users create their own visualizations, \oursystem{} has them select the data they want to visualize from objects ahead of time and then creates the visualizations for them. 

\section{Evaluation}\label{sec:evaluation}
We evaluated the efficacy of \oursystem{}'s framework through a preliminary user study, a comparative study and a series ofusage scenarios. 
\subsection{Preliminary Pair Analytics User Study}
\label{sec:user-study}
User affordances offered by and the development status of a visualization prototype are key factors to steer the design of a user evaluation study~\cite{elmqvist2015patterns}.  In the case of \oursystem{}, we do not intend to validate the scalability or usability of its interface and architecture (see Discussion).  Similarly, we do not evaluate users ability to complete the tasks defined earlier using \oursystem{}. Rather, we found it more appropriate to validate \oursystem{}'s visualization first approach to debugging and the exploration processes that \oursystem{} facilitates. In particular, we wanted to observe the use and utility of global views of program values offered by Anteater in the program exploration process.  Hence, we chose \textit{pair analytics}~\cite{arias2011pair} an appropriate user evaluation protocol.  

% \rf{TODO: Discuss other methods from Elmqvist paper that do not fit our purpose or the state of Anteater at the time.}

Pair analytics offers a ``think-aloud'' protocol that helps generate verbal data by capturing the natural interaction between study participants and the proctor using the visualization interface as a communication anchor.  Using the pair analytics method, a team is formed between a study proctor (or a visualization expert) who helps navigate \oursystem{} and a subject matter expert who drives the exploration/debugging efforts.  

We chose this evaluation over other methods for multiple reasons. First,
this approach allows the subject matter expert to focus less on the nuances of the visualization interface (e.g., interaction types, loading data, etc) and more on exploration and question-answering processes. Other methods require participants to thoroughly learn an entirely new system before completing any tasks. The overhead of learning the nuances of a new system requires a significantly longer study session. Additionally, Anteater is a prototype implementation. Having a proctor to assist in the navigation of the tool provides immediate assistance on how to proceed in the event that a system problem arises in the prototype. Second, a comparative study where experienced programmers complete debugging tasks with Anteater as well as with existing methods not only requires a significant overhead for learning the new system but also must mitigate the bias introduced by participants predisposition towards their current debugging practices. We discuss this more in the next study. 

The exploratory nature of this study combined with the pair analytics protocol allows us to mitigate the bias of a participants predisposition to their current practices and reduce the overhead of learning and using a prototype system while still evaluating the utility of Anteater in exploratory debugging/understanding tasks.

\subsubsection{Methodology} \paragraph{Participants} Participants were recruited from a graduate level ``Principles of Machine Learning" course. All participants are actively involved in computer science research, use Python as their primary programming language, and consider themselves experts in Python. We believe that the debugging and understanding tasks of programs written by graduate students in an upper level machine learning course or their research are comparable to those in real world data analysis programs.

We recruited a total of 5 participants from the class, which had a total of 20 students. However, only 3 of the studies were carried out to completion. We discarded one of these studies because the participant provided a program with a known bug that they thought might be interesting to re-discover with \oursystem{}. While the subject matter expert's program was appropriate for the user study, we thought the prior knowledge of the participant would bias the study's outcome.  As a result, we promoted this program to a usage scenario and discuss it in a later section.  We discarded another study because the programs presented by the participant were not a good fit for the study.  They do, however, highlight some of the limitations of \oursystem{} and are discussed in more detail later.  

\paragraph{Study Session Process}
For each study, we recorded screen capture data along with audio recordings of each interview.  Participants were asked to bring their own program to the study. All participants brought a program that performs some form of data analysis. Allowing them to choose their program helped alleviated some of the mental overhead of the study by not requiring them to learn a new program, along with a new debugging tool. Furthermore, this kept participants in their domain which enabled them to perform more meaningful explorations of their programs with \oursystem{}. 

Participants engaged in two sessions with the proctor. Due to the current policies in place in the U.S. at the time the study was conducted, all sessions were held online rather than in person, as would typically be done. \oursystem{}'s primary developer served as the study proctor to assist participants with navigating the nuances of \oursystem{} and prompting them with questions to describe their exploration process. 

The first session, was a brief meeting to introduce the participant to \oursystem{} and discuss the participants program. The proctor and participants discussed what the participant wanted to see from within their program. After the first session, the proctor ensured that the program was suitable for the study and that the participant could view what they desired by testing it in \oursystem{}. If the program was suitable, participants were asked to meet for a subsequent session. 

In the second session, the proctor walked participants through the various features of \oursystem{}. Afterwards, participants began exploring their programs with \oursystem{}. The study gave participants free reign of their exploration, they were not given specific tasks to accomplish. In doing this, their behavior with \oursystem{} exemplified more precisely how they would use a system like \oursystem{} in their actual program debugging and understanding practices. 

During the second session, the proctor served two primary purposes. First, to mitigate the overhead of learning a new system, the proctor assisted participants in navigating the features of the tool.  Prompted by verbal cues from participants, the proctor would remind participants how to accomplish tasks within \oursystem{}. Second, akin to other pair analytics evaluation studies, the proctor freely asked questions to promote exploratory thinking.  In effect, participants' answering of such questions helped distill internal cognitive processes that were qualitatively analyzed.

For the participants who completed the study, the second session lasted between 60 and 90 minutes.  Approximately the first 30 minutes of each session was spent introducing the subject matter experts to \oursystem{} and getting \oursystem{} set up to run properly on their machines.  

\subsubsection{Results}
\label{sec:eval-results}

From this study, we found that, even in its imperfect prototype state, \oursystem{} was useful to participants for debugging and achieving a better understanding of their programs. All participants were able to learn something new about their program that they previously had not understood. 
For the sake of confidentiality, we cannot give specifics about the programs used by participants.  However, we try to give some context in the form of general concepts found in data analysis programs.

The first participant (P1) knew a bug existed in their program causing it to run incorrectly, but had yet to find it.  With the proctor's guidance, P1 leveraged Anteater to identify and fix the bug (which aligns with {\color{myBlue}G2} above).  Through the use of the timeline plot and the ability to track custom expressions on more complex data structures (which corresponds to T1 - track a variable or expression), P1 found the bug, fixed it, and then verified that the revised program ran properly. During the exploration process, P1 discovered that there was something unusual about the training dataset, denoted as the \textit{whole} dataset, which is split into \textit{left} and \textit{right}, vital to the proper execution of the program.  P1 correctly noticed the problem since ``the right dataset and the [whole] dataset cannot be the same'' even though the scatter plot showed them as identical (T5 - observe relationships between values).  Upon further investigation of the captured values in each dataset, P1 explained that the ``right dataset ... points to [the] class dataset'' which causes them to overwrite the whole training set with only the \textit{right} one.  After modifying the code, a new trace was run and P1 validated the proper behavior of the code.  After being asked if they were ``able to gain new insight into [their] program using Anteater,'' P1 answered that ``the scatter graph and also the tracking values [were] very helpful.''

The other two participants (P2 and P3) presented more open-ended cases. P2 and P3 did not have known bugs, but rather non-trivial data analysis programs whose execution was not fully understood (which corresponds to {\color{myGreen}G3}).  In both cases, the timeline view of certain variables over time was crucial.  P2 heavily relied on the timeline and filtering capabilities of Anteater to verify that their program was converging as expected.  P2 also used the timeline and filtering feature to inspect if their program was reaching the extremes of its search space.  Using the visualizations provided by Anteater, P2 discovered that the program did not search the entire space in one direction and searched beyond the bounds in the other direction.  After completing their exploration, P2 commented that understanding ``why the values are so far off from the [search space] is a good next thing to look at.''

P3 also heavily relied on the timeline view. They used it to understand the behavior of a set of weights in their analysis program.  Before their use of Anteater, P3 had little idea of how the weights behaved throughout their program's execution.  Anteater allowed them to track and visualize the weights over time to see how they evolved as the program ran. After they inspected the weights, the participant commented that ``[Anteater] completely helped [them] understand sort of the underlying domain thing of what was going on with the weights.'' P3 further explained that Anteater was able to show that the program ``is converging on one particular feature as an important weight and the rest [are seen as] super unimportant.'' Through the use of Anteater, P3 was able to understand the behavior of the weights relative to the domain for which the program addressed, specifically through the use of the visualizations. The ability to visualize the variables over time was key to P3 understanding this behavior. 

We believe our observations in this preliminary study provide promising evidence towards the utility of a visualization first approach to exploratory program debugging and understanding. P2 and P3 both performed exploratory tasks for understanding their programs and heavily relied on the global plots of values from within their program and interactions with those plots to improve their understanding of the programs behavior.  In a post exploration interview, all participants indicate that they were able to gain new insight into their programs: P1 by finding their bug and P2 and P3 through understanding the behavior of certain values.  Similarly, all participants expressed that, if a polished and optimized version were available, they would like to use a system like \oursystem{} for future programming tasks. 

% After participants finished their exploration with \oursystem{}, they were asked a set of questions about their experience with it.  The participants were all able to gain new insight into their programs: P1 by finding their bug and P2 and P3 through understanding the behavior of certain values.  When asked what they did not like or find useful, most of their responses were related to usability problems with the current implementation of \oursystem{}, such as confusion about when to right click or left click and when they can alter the source code to have their changes propagate to the back-end. We consider all of their concerns to be minor implementation problems, rather than problems with the underlying system design.  P2 found that for their problem, the histograms were not very useful but could see how such plots could help others.   All participants expressed that, if a polished and optimized version were available, they would like to use a system like \oursystem{} for future problems.  

As mentioned earlier, we discarded one evaluation study, because the programs provided were not ideally suited for the objective of the evaluation.  The participant initially brought a large, machine learning program that took approximately a week to run. This program was not a good fit since we do not aim to study the interaction between trace size and applicability of our approach, but rather the utility of our approach to real Python programmers.  A program that takes a week to run will generate a trace too large to handle by the current implementation of \oursystem{}.  Admittedly, this does limit the generalizability of \oursystem{}, but we consider the study of how to scale a system like \oursystem{} for future work.   The participant then provided several small-scale programs that we were also not ideally fit for the study. The first of the programs was a small multi-threaded program. However, \oursystem{} does not currently support multithreading. The other two programs were linear programs (no loops) with only 20-40 function calls and variable assignments where the code was broken into several small, independent parts.  These programs are similar to those found in an introductory CS course.  This study aimed to use programs similar to those that would be found in a real data analysis setting.  The small-scale programs were simply not sufficiently realistic for the study.   As a result, rather than asking them to provide additional programs, we omitted their case from the study. 

% \subsubsection{Discussion}

% \rf{Something about why this validates our design and what we learned about how people would use this system on their programs}

\subsubsection{Threats to Validity}
Because our sample size was small, the results, while promising, can only be considered preliminary.  We designed the study to keep participants in their domains as much as possible to preserve the ecological validity of the study.  In doing so, we could get deeper insights into users exploration processes and the results would better reflect real world utility of \oursystem{}'s design. However, a full scale study to further validate the design and utility of \oursystem{} needs to be conducted in future work. 

In addition, the pair analytics protocol could potentially introduce bias into the study if the proctor becomes too heavy handed in driving the exploration.  In this study, the goal of the proctor is not to drive the exploration, but rather to aid the user in understanding the nuances of the system.  Their primary role was to observe the participants as they explored their programs, point them to the features of Anteater that would help them answer their questions, and prompt them with additional questions to provoke thought about the findings presented by \oursystem{}.

\subsection{Comparative Evaluation with an IDE}
Our preliminary study aimed at evaluating how programmers use \oursystem{} in their exploratory debugging and understanding tasks. We found that the approach of providing global views of program values at the forefront was useful for our participants in completing exploratory program understanding tasks. In a second study, we attempt to compare how participants debug with \oursystem{} vs a more traditional IDE debugger. In this study, we asked people to complete two debugging tasks, one with \oursystem{} and one with an IDE.  We then prompted them with questions about their experiences.  The goal of this study was not to evaluate how efficiently  or accurately participants debugged programs. Rather, we aimed to evaluate how people interact with each system and contrast their experiences. However, over the course of the study we encountered several challenges to conducting an evaluation of this type. These will be discussed more in the following sections.

% \rf{For \oursystem{}, the study focused on problems that primarily required participants to use the the global views of variable values rather than evaluating all features of the system. Participants were instructed on how to use the full system, but we were unable to create debugging tasks that highlighted all major features of the tool. }

\subsubsection{Methodology}
Using recent technology that brings the Python library into the browser~\cite{pyodide}, we developed a version of \oursystem{} that runs entirely in the browser and does not require a python server back-end as in the prior version. Thus, in contrast to our preliminary study, participants were able to complete this study entirely in the browser, without any assistance from a proctor or system expert. Participants were asked to complete 2 debugging tasks: one with \oursystem{} and one with an online Python IDE.   

\paragraph{Participants} Throughout the course of the study, we conducted two rounds of recruiting participants.  

The first round of recruiting was conducted by advertising the study to relevant groups of research and data analysis professionals. As an incentive for completing the study, participants were entered into a drawing for a \$100 Amazon giftcard. We chose this method of recruitment because we believed that these groups of professionals would have the experience to provide focused and meaningful feedback. We kept the study open for 4 weeks. However, we were only able to recruit 4 participants in this first round. As a result we  conducted a second round of recruitment. 

In the second round of recruiting, we recruited participants through the recruitment service Prolific. To ensure that our participants had the proper experience, Prolifc allowed us to specify that participants must have programming experience.  Additionally, in the description of the study, we specified that they must have Python programming experience. However, despite this specification, several participants indicated that they were novices with Python, which may have impacted their ability to complete the given debugging tasks. These participants tended to provide the least meaningful responses.  Through prolific, we recruited 9 participants. Participants were paid \$10/hour to complete the study. 

Of the 13 participants, 4 rated themselves as Python novices (have never used Python before or are currently learning it), 6 as intermediate (use Python sometimes, but not as a primary language), and 4 as experts (use Python regularly as a primary language). 

Participants were asked to provide the purpose of their primary programming activities. 4 of the participants indicated that they primarily program for coursework, 7 indicated that the program for software development and 2 indicated that they program for data science/analysis.  

Participants were also asked about their current debugging practices. 8 participants indicated that print statements were part of their debugging process, 8 use breakpoint debuggers, and 6 participants use a mix of the two. 3 participants did not provide descriptive responses.

\paragraph{Study Setup} During recruitment, we gave participants a link to a webpage describing the study purpose and format. If participants chose to start the study, they were taken to a Google form to where they were they were shown an instructional video on how to use \oursystem{} and then given two debugging tasks to complete, one with \oursystem{} and one with an IDE. Participants were asked to debug the program either until they found the bug or until 10 minutes lapsed. We wanted participants to try to debug the program but, in the event that they could not find the bug, we did not want to ask them to spend more than 10 minutes trying. 
After each debugging task, participants were asked a series of questions about their experiences.  Upon completing both tasks, they were asked a series of questions to compare their experiences and evaluate the design of \oursystem{}. 

\paragraph{Debugging Tasks} The study asked participants to complete two debugging tasks. Both tasks involved computing the number of polling places per capita for every state in the US from a dataset containing the population and number of polling places for each county in the country.  One of the debugging tasks (D1) generated erroneous (negative) values due to an off by one error when indexing a list that separated each state in the data by inserting a value of -999.  The second debugging task (D1) generated similarly erroneous values due to an unclean data file that represented missing values with -999. Participants were presented the two tasks in random order and each task was randomly paired with either \oursystem{} or the IDE.

\subsubsection{Results}
Overall, the reception of \oursystem{} was generally positive, especially since this was participants first exposure to the system. 

Of the 13 participants, using either tool, only 5 were able to confidently find  D1 (4 were unsure and 5 did not believe they found it) and only 2 were able to confidently find D2 (7 were unsure and 4 did not believe they found it). This indicates that the bugs we introduced, while seemingly simple, were likely  too complex for this type of study.  This exemplifies the first challenge we encountered in this study: introducing sufficiently small but realistic and meaningful bugs that are identifiable in a short period of time. We carefully developed these bugs and tested them in a pilot study to ensure that they were not too complex. However, despite this, it seems our bugs were still to complex.

Additionally, several participants cited that the recommended time limit of 10 minutes for each task was not enough to learn \oursystem{} and complete the debugging task. One participant noted \textit{``I felt like anteater was aiming towards a feature I would find useful, but also only having 10 minutes of time with it, it certainly wasn't effortless to figure out and I suspect I missed some possible chances to take better advantage of it.''}. This highlights the second challenge we encountered: overcoming the overhead of teaching participants an entirely new system while still keeping the study a reasonable length. Most participants took at least an hour to complete the study with the 10 minute time limit. Given the offered compensation, we did not feel that would could ask for more time than that. However, it seems that this was not enough time for participants to learn the nuances of the system.

\begin{figure}[t]
    \centering
    \includegraphics[width=\linewidth]{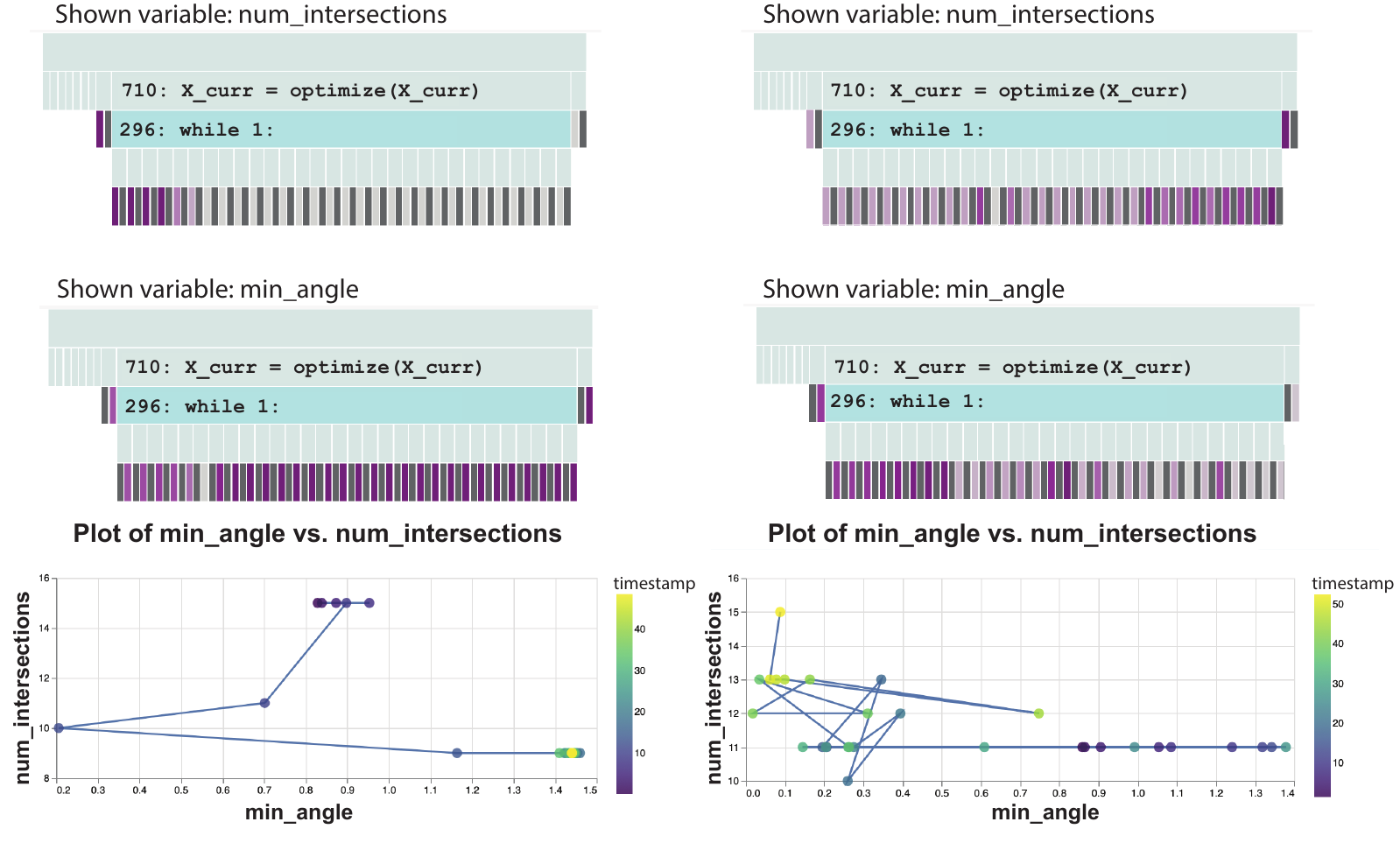}
    \caption{Using \oursystem{} to compare two runs of gradient descent that should maximize the minimum crossing angle while minimizing edge crossings. The generalized context trees in (A) show that the number of intersections rapidly decreases (the color changes from dark purple to white) while the minimum angle increases. The scatterplot shows that the descent spends its first few steps at a bad solution and takes approximately three big steps before converging on a good solution.  In contrast, in (B) the number of intersections increases throughout the descent while the minimum angle decreases.  The scatterplot shows that, in general, as the number of intersections grows, the minimum angle shrinks and lands at a bad solution.}
    \label{fig:vanillaComp}
\end{figure}
\begin{figure*}[t]
    \centering
    \includegraphics[width=\linewidth]{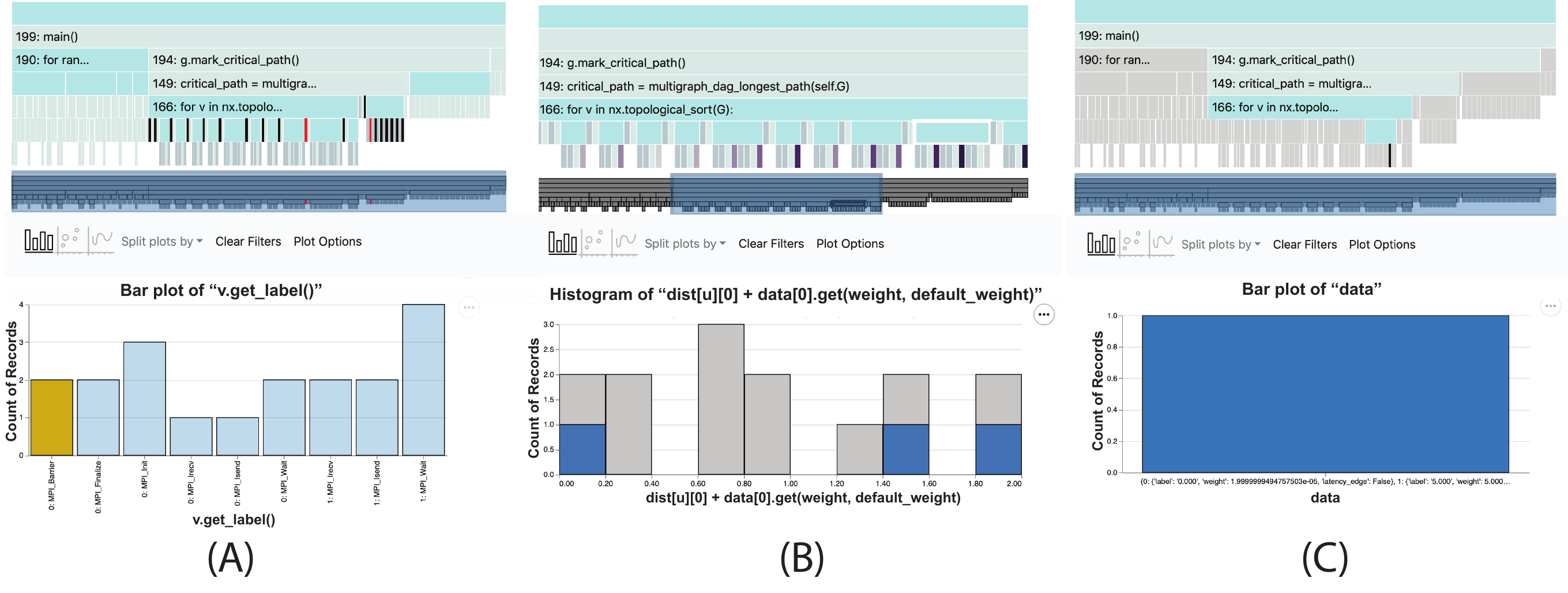}
    \caption{(A) shows how we can use the bar plot to find where the barrier node occurs in the execution plot.  (B) shows the weights of the paths to the barrier node calculated in the program. We see that none of them are 5, as we expect from looking at the MPI call graph.  (C) shows the data used to calculate the path weight that should include the edge of weight 5.  We see that the data does include this edge but the algorithm does not look at it. For more details, refer to the supplemental video. }
    \label{fig:MPI}
\end{figure*}

After completing their debugging tasks, around half of the participants (7 out of 13) indicated that they would prefer a production ready version of a system like \oursystem{} over an IDE. One participant cited\textit{ ``Anteater seems like it would be more effective for the targeted debugging style that I tend to follow, i.e. focus on a few specific variables or expressions, and investigate them deeply (and being able to do it visually isn't something supported in anything I've seen!), as opposed to the way the IDE just gives me cluttered lists of numbers for everything, even if I already know they're not important.''} Another participant cite that Anteater \textit{``removes the careless of skipping the thing that we wanted to find (odd values)''} that happens when stepping in IDE debuggers.

Of the 6 who preferred traditional IDE's, 4 of them cited that they preferred the IDE and print statement debugging simply because they are accustomed to it. One participant cited \textit{``I'm just more used to the traditional IDE, it's debugging paradigm makes sense in my head. Anteater introduces a whole new paradigm of debugging, which I could imagine to be useful, but I'm just not used to it''}.  Another stated that Anteater \textit{``provides a lot more information about what's happening with the problem, but I think it does not replace the phase of using convenient prints and proper testing to check what's exactly the problem.''}. This exemplifies one of the largest challenges that we encountered during this evaluation: overcoming experienced programmers predisposition to their current debugging practices. One participant strongly stated that \textit{``I find prints the way of God''}.
Several participants felt that \oursystem{} would be useful but they did not want to change their current debugging practices. However, two of these participants commented that they think the features of \oursystem{} would be useful if integrated into an IDE.  One participant stated `\textit{`The best combination would be to integrate Anteater in a classical IDE in order to have the best of both worlds''}. We found this a very encouraging comment for validating \oursystem{}'s visualization first approach to debugging and will explore this possibility in future work.

% \rf{For D1, only 1 (out of 4) participant was able to find the bug using \oursystem{} while 4 (out of 9) participants were able to find it with the IDE. Note, due to the small sample size, the randomizer disproportionately gave participants D1 to complete in the IDE and D2 to complete with \oursystem{}. }

% (D1) Anteater:yes 1, unsure 1,  no 2, , IDE: yes 4, unsure 2, no 3,
% (D2) Anteater :yes , unsure 6, no 3, IDE: yes 2, unsure 1, no 1, 

% \rf{We were initially hesitant to conduct a study of this type because we believed that it would be difficult to overcome participants predisposition towards traditional methods. }

\subsubsection{Threats to Validity}
While the first round of targeted recruitment produced more meaningful results, there is the potential for bias due to the participants potential familiarity with the authors. Although the study was anonymous, we recruited participants from groups with which the authors are affiliated. It is possible that participants were more generous with their responses than they would have been if they did not know the authors but we have no way of detecting this. 

In the second round of recruiting, we were unable to properly screen participants through Prolific.  In order to maintain the validity of the study, we did not modify the protocol when extending the recruitment to Prolific. However, it seems that the study would have benefited from additional efforts to screen participants Python and professional experience. In Prolific, we were only able to specify that participants have programming experience. However, Python experience was important for the study. One participant was unable to complete either task, citing that both tools were not accessible, possibly due to the fact that they were a novice in Python.

\subsection{Usage Scenarios}
\label{sec:casestudies}
Here, we present several, real-world scenarios, showcasing how \oursystem{} derives insight into debugging and program understanding. These scenarios were developed on real programs through the author's debugging efforts using \oursystem{}. 
\subsubsection{Gradient Descent}

The first usage scenario we present inspects a program performing gradient descent.  This program was collected from a question on Stack Overflow~\cite{stackOverflow}. The programmer struggled to figure out why the resulting values of the variables ``$x$'' and ``$x1$'' were NaNs.  We will walk through how to use \oursystem{} to understand the bug and correct it.  

First, we run the program with \oursystem{} to track one of the misbehaving variables, ``$x$."  Fig.\ \ref{fig:gdOverview}-A shows the resulting GCT and histogram.  The histogram shows that much of the descent generates NaNs (the green bar).

As a natural next step, we look at these values over time. We switch the plot type to ``scatterplot'' which shows a plot of the variable ``$x$'' over time, shown in Fig.\ \ref{fig:gdOverview}-B.  Now, we clearly see that the value of ``$x$'' stays around zero, before becoming a very small negative, then going to infinity after which it reaches the NaNs. However,  something strange happens where the value stays around zero and then suddenly becomes a very small negative.  To investigate this, we filter the values to show only those points staying close to zero. We also switch to a symmetric log scale because we suspect that the values may not actually lie that close to zero. 
Fig.\ \ref{fig:gdOverview}-C shows the resulting visualizations.  We see that the value oscillates between increasingly large positives and negatives until it reaches infinity. 

Now that we know the problem, we try to fix it. The oscillating values suggest that the gradient is exploding due to a training rate that is too large.  In Fig.\ \ref{fig:gdOverview2}, after lowering the training rate and re-running the trace, the value quickly converges, as expected. 

Using \oursystem{}, we quickly and easily track the variable ``$x$" and see its behavior throughout the execution.  In a traditional debugger, detecting this behavior requires stepping through several iterations to view the values.  After lowering the training rate, we repeat this process to determine if that fixed the problem.  This involves significantly more interaction with the debugger than when using \oursystem{}.

\begin{table*}[htbp]
    \small\sf\centering
    \begin{tabular}{|c|c|c|c|c|c|c|c|}
        \hline
        \multirow{3}{*}{\centering Program} & \multirow{3}{1.4cm}{\centering \# Tracked Variables} &  \multirow{3}{*}{\centering \# Lines} &  \multirow{3}{1.5cm}{\centering \# Recorded Calls} &  \multirow{3}{1.7cm}{\centering \# Recorded Assignments}  &  \multirow{3}{1.5cm}{\centering Original Execution Time (s)} &  \multirow{3}{1.7cm}{\centering Instrumented Execution Time (s)} &  \multirow{3}{1cm}{\centering Trace Size (MB)}  \\
        & &&&&&&\\
        & &&&&&&\\
        \hline
        Gradient Descent & 5 & 36 & 402 & 801 & 0.0041 & 0.0691 & 0.3098\\
        Longest Weighted Path & 3 & 199 & 149 & 55 & 0.0023 & 0.0343& 0.2351\\
        Recursive Fibonacci & 1 & 11 & 150,049 & 75,024 & 0.0292 & 5.8952 & 116.2\\
        \hline

    \end{tabular}
    \caption{Trace information and execution impact for traces of three of the programs discussed in this paper. The trace size and performance impact depends on the amount of information recorded in the trace}
    \label{tab:performance}
\end{table*}
\subsubsection{Graph Edge Crossing Angle Maximization}
In this usage scenario, we investigate a program that tries to balance the number of edge crossings in a graph with the size of the minimum crossing angle.  The program searches for the layout that minimizes the number of edge crossings while maximizing the size of the minimum crossing angle.  In this usage scenario, we inspect the stability of the gradient descent method on this problem. 

To inspect the stability, we ran the gradient descent multiple times, tracking the minimum angle and number of intersections at each iteration of the gradient descent.  We found that in most cases, the gradient descent returns a good solution, as demonstrated in Fig.\ \ref{fig:vanillaComp}-A, where it immediately begins moving toward a good solution and never turns back. However, instances occur, as shown in Fig.\ \ref{fig:vanillaComp}-B, where the gradient descent starts moving towards a bad solution, and never recovers.  Therefore, we can conclude that although the majority of the time it produces a good solution, this method suffers from stability issues.  

\subsubsection{Longest Weighted Path Calculation}

This usage scenario was presented to us by a prospective participant in the user study. While the program was not a good fit for the study, because the participant already knew where the bug was, it presents a good example of the utility of \oursystem{} on real problems.  This program aims to find the critical path, i.e., the longest weighted path, from the ``Init'' to ``Finalize'' nodes in an MPI call graph.  It uses the networkx library to build a multiDAG and calculate the longest (weighted) path.  We were given this program with the knowledge that this bug existed and which methods were affected but no other information on how to fix it.  We then found and fixed the bug using only \oursystem{}.  Below, we explain how we found the bug. 

To begin, we loaded the program and data files into \oursystem{}. We know from inspecting the test graph manually that the algorithm overlooks one of the edges (of weight 5) from the ``Init'' node into the ``Barrier'' node. To build the longest path, the algorithm topologically sorts the nodes and iterates over them.  For each node, it iterates over all of the predecessor nodes.  To find the bug, we first need to find where the barrier node occurs.  We do this by collecting the node label in each iteration.  Inspecting the node labels in the bar plot, as shown in Fig~\ref{fig:MPI}-A, shows us the point in the execution tree where the loop reaches the ``barrier'' node.  Once we find the barrier node, we select it to view the other values in that specific iteration. We then switch variables to look at the path weights for each predecessor, as shown in Fig.\ \ref{fig:MPI}-B.  We see (from the x-axis of the barplot) that none of the path weights reach 5, which indicates that the algorithm misses the edge of weight 5 into the ``Barrier'' node.  Next, we look at the data used to calculate the path weights.  We notice that one of the ``Init'' predecessors has two weights associated with it, as shown in the filtered bar in Fig.\ \ref{fig:MPI}-C.  There are two keys in the dictionary, one for each edge from ``Init'' to ``Barrier''. Looking back at the algorithm, we see that it only looks at the first key which causes it to miss the edge of weight 5 and report an incorrect longest weighted path. To fix this, we simply find the edge with the highest weight over all of the edges from the predecessor to current node.  

\section{Discussion and Limitations}
\label{sec:discussion}

\paragraph{Omnicode vs. \oursystem{}}
While Omnicode and \oursystem{} both intend to help programmers debug and understand their programs, the two systems differ
in their target audience.  Omnicode aims to help novice users create  mental-models to reason about their program's execution and debug unexpected behavior.  The size and complexity of programs it needs to support for this audience is quite small. Thus, Omnicode only supports programs of around 10 variables and 100 execution steps. \oursystem{} aims to help programmers in general. Therefore it needs to support different types of programs.  

While \oursystem{} cannot support large scale software-systems as they produce an unmanageable amount of data, it can support much larger programs than those written by novices, such as those programs written by data scientists. \oursystem{}'s ability to support a program largely depends on the number of function calls and variable assignments that are recorded in the trace. We do not know the exact limits of these parameters (as they are very interdependent) but know that \oursystem{} certainly supports programs with up to 225,000 function calls and variable assignments. Each function calls and variable assignment is comparable an execution step as described by Omnicode. Thus, in contrast to Omnicode which supports programs of around 100 execution steps, \oursystem{} can support programs of at least 225,000 comparable steps. 

Most of the differences between Omnicode and \oursystem{} stem from the fact that they are geared toward different audiences. Omnicode supports a live programming environment because it targets small programs whereas a static environment makes more sense for \oursystem{}. Similarly, Omnicode tracks every variable in the program which is infeasible for the larger programs \oursystem{} supports.

\paragraph{Goal 3} As mentioned earlier, {\color{myGreen}G3}'s serves as a catch all for debugging tasks that do not fit into the first two goals. We acknowledge that a different, more specific, version of {\color{myGreen}G3} may exist that, when evaluated, would allow us to learn more specific information about how programmers use \oursystem{}. However, keeping this goal general allowed us to support any exploratory tasks presented to us by potential participants in the preliminary study. This study focused more on observing how people would use \oursystem{} to validate its design rather than evaluating their ability to complete tasks with \oursystem{}.  

\paragraph{Choosing what to track} 
As stated earlier, \oursystem{} only collects the variables and expressions that the programmer specifies. We explicitly chose to do this because it reduces the amount of unnecessary information presented to the programmer. However, in some cases, such as when a programmer does not quite know what variable contains the bug, people may want suggestions of variables to inspect or they may want to inspect all variables. The problem of automatically suggesting variables and efficiently tracing all program variables remains for future work. One possible approach could be to still have the programmer specify variables, but then automatically collect all values that the variables depend on. 

\paragraph{Limitations}
\oursystem{} will not scale to programs the generate large traces.  Such programs typically make many calls or assign to tracked variables many times.  In these programs, the traces become too large and the visualizations unreadable. Table~\ref{tab:performance} shows the performance impact for several programs traced with \oursystem{}. Research exists on collecting the entire trace of large programs~\cite{Pothier2017}; future work is needed to evaluate if \oursystem{} works well with this method. We note that our visualizations operate on relational data, and there
is a growing number of techniques to support interactive visualizations on large relational datasets~\cite{godfrey2016interactive,moritz2017trust}. A full investigation of their impact on program visualization, however, is out of present scope.
In addition, \oursystem{} works best with numerical data and has limited support for other datatypes.  While it can present numbers, strings, and booleans, it does not support compound objects directly. Information about variables of these datatypes can still be visualized through the use of custom expressions, but we leave first-class support for more datatypes for future work. 
Finally, \oursystem{} assumes a sequential programming model and does not support parallel programs.  Work exists in automatic tracing of parallel programs in the traditional sense (without values) but applying and extending these traces to \oursystem{} is left for future work.

\bibliographystyle{SageV}

\bibliography{paper.bib}

\end{document}

\title{Supplemental Material for ``Anteater: Interactive Visualization of Program Execution Values in Context''}
\author{}
\date{}
\maketitle

Our supplemental material can be found at {\color{blue}\url{https://osf.io/j4byp/?view\_only=bffbc752dd7943108221baf31dd163e6}}

The supplemental material includes:
\begin{enumerate}
    \item AnteaterDemo.mp4 - A video demonstrating how Anteater was used in the Gradient Descent program discussed in the case studies section of the paper. 
    \item AnteaterMPIExample.mp4 - A video demonstrating how Anteater was used in the Longest Weighted Path Calculation example in the case studies section of the paper.
    \item AnteaterDemo.zip - A demo version of Anteater.  This zip file contains a front-end only version of Anteater with several pre-loaded programs and traces. The provided programs include 3 of the 4 programs discussed in the paper (Fibonacci, Gradient Descent, and Longest Weighted Path) along with one additional program that explores different data types.  The file also includes instructions for running the demo. 
\end{enumerate}